\documentclass[12pt]{iopart}

\usepackage{iopams}

\usepackage{soul,wasysym}
\usepackage{tabularx, booktabs,ragged2e,hyperref}
\usepackage{bbold}
\usepackage[svgnames,table]{xcolor}

\usepackage{blindtext}
\usepackage{graphicx}
\usepackage{verbatim}
\usepackage{color}
\usepackage{subfigure}
\usepackage{placeins}
\usepackage{bm}
\usepackage{bbm}
\usepackage{amsfonts}
\usepackage{amsbsy}
\usepackage{amssymb}
\usepackage[nameinlink,noabbrev,capitalize]{cleveref}

\crefname{appendix}{}{}
\usepackage[normalem]{ulem}
\useunder{\uline}{\ul}{}
\usepackage{adjustbox,lipsum}
\usepackage{float}
\usepackage{array}
\usepackage{multirow}

\makeatletter
\long\def\@makefntext#1{\parindent 1em\noindent 
 \makebox[1em][l]{\footnotesize\rm$\m@th{\arabic{footnote}}$}%
 \footnotesize\rm #1}
\def\@makefnmark{\hbox{${\arabic{footnote}}\m@th$}}
\def\@thefnmark{\arabic{footnote}}
\makeatother
\newcommand{\R}{\mathbb{R}}

\newcolumntype{Y}{>{\centering\arraybackslash}X}
\newcolumntype{C}{>{\centering\arraybackslash}X} 

\newcommand{\beq}{\begin{equation}}
\newcommand{\eeq}{\end{equation}}
\newcommand{\bqa}{\begin{eqnarray}}
\newcommand{\eqa}{\end{eqnarray}}
\newcommand{\bal}{\begin{equation}\begin{aligned}}
\newcommand{\eal}{\end{aligned}\end{equation}}

\newcommand{\erf}[1]{Eq.~(\ref{#1})}

\newcommand{\erfs}[2]{Eqs.~(\ref{#1})--(\ref{#2})}

\newcommand{\srf}[1]{Sec.~\ref{#1}} 
\newcommand{\crf}[1]{{Ref.}~\cite{#1}} 
\newcommand{\trf}[1]{Table ~\ref{#1}}

\newcommand{\frf}[1]{Fig.~\ref{#1}}

\newcommand{\dg}{^\dagger}

\newcommand{\bra}[1]{\left\langle #1\right|}
\newcommand{\ket}[1]{\left|#1\right\rangle}
\newcommand{\braket}[2]{\left\langle #1|#2\right\rangle}
\newcommand{\ketbra}[2]{| #1\rangle\langle #2|}

\newcommand{\blk}{\color{black}}

\newcommand\stPW{\bgroup\markoverwith{\textcolor{red}{\rule[0.5ex]{2pt}{0.4pt}}}\ULon}

\newcommand{\ro}[1]{\left( {#1} \right)}

\newcommand{\no}[1]{}

\begin{document}

\title{Symmetries and physically realizable ensembles for open quantum systems}
\date{\today}

%

\author{Prahlad Warszawski\textsuperscript{1,3} and Howard M Wiseman\textsuperscript{2}}

\address{\textsuperscript{1}Centre of Excellence in Engineered Quantum Systems (Australian Research Council), School of Physics, The University of Sydney, Sydney, New South Wales 2006, Australia}
\address{\textsuperscript{2} Centre for Quantum Computation and Communication Technology (Australian Research Council), 
Centre for Quantum Dynamics, Griffith University, Brisbane, Queensland 4111, Australia}
\address{\textsuperscript{3} Author to whom correspondence should be addressed}
\ead{prahladw@gmail.com}


\begin{abstract}
A $D$-dimensional Markovian open quantum system will undergo 
stochastic evolution which preserves pure states, 
if one monitors without loss of information the bath to which it is coupled. 
If a finite ensemble of pure states satisfies a particular set of constraint equations 
then it is possible to perform the monitoring in such a way that the (discontinuous) trajectory of the conditioned system state is, at all long times, restricted to those pure states. Finding these physically realizable ensembles (PREs) is typically very difficult, even numerically, when the system dimension is larger than 2.  In this paper, we develop symmetry-based techniques that potentially greatly reduce the difficulty of finding a subset of all possible PREs. The two dynamical symmetries considered are an invariant subspace and a Wigner symmetry.  An analysis of previously known PREs using the developed techniques provides us with new insights and lays the foundation for future studies of higher dimensional systems.

\end{abstract}


\maketitle
\section{Introduction}

Given an open quantum system obeying a Markovian master equation (ME), an experimentalist  typically has much choice as to the way in which the environment, with which the system interacts, is monitored.  Given this experimental freedom, one can ask the question of whether the system state can be made, in the long time limit, to jump between a finite number of pure states, with static dynamics between jumps.  That is, can the evolution of the system state be made to follow a Bohr-Einstein model of jumps between quantum states that are, conditional upon measurement results, stationary (although not necessarily the eigenstates of some Hamiltonian)? If it can, how many such states, at a minimum, are required? Alternatively, what is the minimum Shannon entropy of the ensemble?  Questions of this type~\cite{karasik2011many} may offer novel insights into the nature of open quantum system dynamics and the quantum--classical epistemic distinction~\cite{2001quant.ph..1012H,Gu2012,monrasPub}.  The observation of single quantum trajectories has now been realized in the laboratory~\cite{vijay2011observation,murch2013observing,PhysRevX.6.041052,PhysRevX.6.011002,minev2018catch}, with ever increasing detection efficiencies~\cite{PhysRevX.9.011004}. Additionally, there is strong theoretical evidence for the stability of the particular class of quantum trajectories under consideration~\cite{karasik2011tracking}. Consequently, our questions are not exclusively of theoretical concern.  
\blk

It might have been thought that since there is an infinity of ways to write a state matrix as an ensemble of, in general, non-orthogonal pure states, there would also be an equivalent diversity of ensembles that are realizable via the application of an appropriate measurement scheme.  However, this has been shown to be incorrect~\cite{wiseman2001inequivalence}: for some ensembles there is no way that the experimentalist can know at all (long) times that the system is in one of the states belonging to that ensemble.  Ensembles that are realizable are known as {\it physically realizable ensembles} (PREs).  Such ensembles support an {\it ignorance interpretation} in the sense that an individual who is not privy to the experimental results, produced during the ensembles realization, could claim that the system really is in one of the pure states belonging to the PRE. This provides an approach to understanding the emergence of classical behavior in open quantum systems, an important, and ongoing, topic of study~\cite{RevModPhys.75.715,frowis2018macroscopic}.  

Given that not all ensembles are experimentally realizable, the existence of finite ($K$-element) PREs, for finite dimensional open quantum systems, is non-trivial and perhaps even surprising --- typically, a quantum trajectory will traverse a continuum of states during periods of monitoring that don't involve a detection event.  In this sense, finite PREs are {\it special}: they highlight the power of measurement in inducing certain types of quantum trajectories.  The perspective afforded by $K$-element PREs may allow unique insights into the dynamics of open quantum systems.  Note that this phenomenon has no classical analogue, as for a discrete classical stochastic system the set of pure states is purely kinematics, independent of dynamics.  An additional, but related, motivation for the study of PREs is the desire to minimize the
resources (memory) required to track an open quantum system. The central consideration, which was raised in Refs. \cite{karasik2011many,karasik2011tracking}, is that if a $K$-state PRE can be found for a given ME, then it is possible to track that same open quantum system with a $K$-state classical memory.  Given that it is provable that quantum systems of dimension $D=2$ can always be monitored using a $2$-state classical memory~\cite{karasik2011many}, investigation of PREs in higher dimension must be conducted in order to test whether there is generically a gap between the size of the smallest ensemble and $D$.  An important purpose of the current paper is to facilitate this investigation.

\blk

In Ref.~\cite{wiseman2001inequivalence}, a set of polynomial constraint equations were discovered that govern the existence (or otherwise) of PREs.  A solution set of the constraints would describe all possible PREs for a given ME.  These constraints will be detailed in the next section, but in our introductory remarks we wish to discuss their generic solubility.    Unfortunately, solving a set of nonlinear polynomial equations typically becomes exponentially more difficult as the number of equations and variables increases.  In fact, the problem is known to fall into the NP-complete complexity class~\cite{courtois2000efficient}.  This difficulty is acute in the case of the PRE constraint equations in question, as the number of constraints in the set scales as $KD^{2}$, where $K$ is the number of pure states in the ensemble and $D$ is the system dimension.  Thus far, a full description of PREs for a given ME --- via analytic or numeric means --- only exists in the literature for $K=D=2$~\cite{karasik2011many}.  Tied to the difficulty of finding PREs, for the cases that have been the focus of research thus far (typically searches for the minimal, or close to minimal, sized ensembles), is the fact that the polynomial systems describing their existence are highly constraining, in the sense that PREs occupy a volume of measure zero of the parameter space.  In other words PREs are rare; despite all the measurement freedom available to an experimentalist, almost all small-$K$ ensembles are not realizable.  This essentially precludes the technique of guessing a potential PRE and then trying to design a measurement scheme to realize it: the parameter space is enormous and presents a much more difficult search than solving the constraints of \cite{wiseman2001inequivalence} directly.
It is perhaps possible that, via detailed numerical study of large-$K$ ensembles, 
one may find sets of PREs that occupy a finite volume of the parameter space.  In this paper, we focus on developing theoretical techniques that are applicable to arbitrary sized PREs.  

\blk

Here, we introduce theoretical methods that can simplify the task of finding example PREs for a ME.  This is important as it will make tractable the study of PREs for a broader range of MEs and ensembles, in particular MEs for systems of larger dimension, $D>2$, and PREs of greater than minimal size.  The primary tool that we introduce is that of symmetry; many physical systems of interest possess symmetries and even generic systems possess some symmetry (by our definitions, as will be made clear in the appropriate basis).
In Ref.~\cite{karasik2011tracking}, it was stated that PREs of smaller $K$ (than otherwise expected) could perhaps be discovered by exploiting additional structure within the ME: in this paper we develop such ideas in considerable detail, with the conclusion that there is, indeed, much to be gained from an analysis that utilizes system symmetries.  Due to the, still considerable, computational difficulty of finding PREs in $D>2$, we defer these searches to future work.  In the current paper, we lay the groundwork with an exploration of previously analyzed MEs and PREs from a new perspective that gives increased understanding of, and experience with, our new techniques.

Two types of symmetry are explored, followed by a consideration of their joint application.  Firstly, we consider an `invariant subspace symmetry', by which we mean that the dynamics of the system state are contained to within some region of the space of density matrices, given an initialization within that space.  Our definition contrasts with other well-known invariant subspace definitions~\cite{4639467} and allows us maintain the dimensionality, $D$, of the system but, nevertheless, decrease the size of the space that is searched for the presence of PREs.  This reduces the computational task of identifying a PRE.  Secondly, we consider `Wigner symmetries', which are those transformations that preserve the inner product of Hilbert space 
rays~\cite{wigner1959group}.  The consequences of such a symmetry existing are explored for PREs: we find that new PREs can be generated via the application of a Wigner symmetry to an already known PRE and, also, that some fraction of the constraint equations become redundant given a PRE that is Wigner symmetric (a concept that will be defined).  Both of these discoveries will make a study of PREs possible in some situations where their finding was previously intractable.  Detailed case studies for these symmetries are made for two single-qubit MEs: the resonance-fluorescence ME and the thermal equilibrium (absorption and emission) ME.

The organization of this paper is as follows.  Firstly, we provide a more mathematical description of PREs, in \srf{PREintro}, in order to set the grounds for our in-depth study.  In \srf{feas}, we motivate the need for new techniques to find PREs, based on a discussion of the inherent computational difficulty.
Next, in \srf{invSubspaceSec}, we consider the invariant subspace symmetry, followed by a discussion of its utility and qubit examples. Then, in \srf{uniSymm}, we consider Wigner symmetries and how they can generate both new PREs and be applicable to an individual PRE.  Once again we highlight with qubit examples.  In \srf{combo}, we consider how both these symmetries may co-exist.  This is then followed by a conclusion and discussion in \srf{conc}.

\section{PRE preliminaries}
\label{PREintro}
This paper is concerned with autonomous Markovian open quantum systems of finite dimension $D$.  In the absence of measurement (or if the measurement results are ignored) the dynamics of such systems are governed by a Lindblad-form master equation (ME) for the density matrix\blk~\cite{WisMil10}: 
\beq \label{me1}
\dot{\rho} = {\cal L}\rho \equiv -i[   \hat{H}_{\rm eff}\rho   -  \rho \hat{H}_{\rm eff}\dg] 
+ \sum_{l=1}^L  \hat{c}_l \rho  \hat{c}\dg_l,
\eeq
where $ \hat{H}_{\rm eff} \equiv \hat{H} - i\sum_l   \hat{c}\dg_l  \hat{c}_l/2$ and $\hat{H}$ is the Hermitian Hamiltonian.  Without loss of generality we take the Lindblad operators $\{ \hat{c}_l\}$ as linearly independent and traceless but otherwise arbitrary, which leads to a bound of $L\leq D^2-1$~\cite{doi:10.1063/1.522979}.  The separation of \erf{me1} into terms that involve $ \hat{H}_{\rm eff}$ and those comprising $\hat{c}_l \rho  \hat{c}\dg_l$ can be associated with a purity-preserving \blk unraveling of the ME as would arise from perfectly efficient \blk  monitoring of the decoherence channels, $l$.  Assuming an initially pure system state,  if \blk a detection in channel $l$ is observed at time $t$ then the system jumps from the pre-jump state $\ket{\psi(t^{-})}$ to the post-jump state $\ket{\psi(t)}\propto \hat{c}_l \ket{\psi(t^{-})}$.  After the jump, the quantum state evolves under the no-jump evolution operator $ \hat{H}_{\rm eff}$ and will {\it not} remain stationary unless it happens to be an eigenstate of $ \hat{H}_{\rm eff}$.  

We limit the MEs under consideration to those that produce an impure, but unique, steady-state density matrix, $\rho_{\rm ss}$, of rank $D$, defined by ${\cal L}\rho_{\rm ss}=0$. It is always possible to write $\rho_{\rm ss}$ as a mixture of pure states, but because the pure states in this decomposition are not necessarily orthogonal there are an infinite number of ways of doing so.  However, only a subset of these ensembles --- termed physically realizable ensembles (PREs) --- can be realized at all sufficiently long times as the states conditioned on the outcomes of some 
monitoring of the decoherence channels. This 
might seem obvious due to the requirement that the ensemble members be eigenstates of the no-jump evolution operator $\hat{H}_{\rm eff}$.   However, that would be a premature judgement, as we need to consider {\em adaptive} detection, for which the operator $\hat{H}_{\rm eff}$ can change, even while \erf{me1} remains fixed. This will be explained fully below. 

More than one PRE can be achievable (in separate experiments of course) for a given ME due to the invariance of \erf{me1} under the following joint transformations~\cite{PhysRevA.49.2133,karasik2011many}
\bqa
 \hat{c}_{l}&\rightarrow&\left\{ \hat{c}^{\prime}_{m}  = \sum_{l=1}^{L} S_{ml} \hat{c}_l + \beta_m\right\}  \label{jumpOps}\\
\hat{H}&\rightarrow&\left\{ \hat{H}^{\prime}= \hat{H} - \frac{i}{2}\sum_{m=1}^{M}  (\beta^{*}_m   \hat{c}^{\prime}_m - \beta_m  \hat{c}^{\prime\dagger}_m{}) \right\},
 \label{noJumpOps}
\eqa
where, with $M\geq L$, $\pmb{\beta}$ is an arbitrary complex $M$-vector and ${\bf S}$ is an arbitrary $M\times L$ \blk
semi-unitary  matrix; that is, \blk 
$\sum_{m=1}^{M} S^{*}_{ml}S_{ml'} = \delta_{l,l'}$. 
By unraveling \erf{me1} with $\{\hat{c}^{\prime}_{m}\}$ as the jump and 
$\hat{H}^{\prime}_{\rm eff} =  \hat{H}^{\prime} - i \sum_{m=1}^M \hat{c}^{\prime}_m{} \dg \hat{c}^{\prime}_m /2$ as the no-jump operators, different $\pmb{\beta}$ and ${\bf S}$ thus correspond to different measurement schemes~\cite{CarQTraj,WisMil10}. 
It is important to note that the implied number of detectors (equal to $M$) can be greater than
the number of Lindblad operators, $L$, describing the ME.  

The above flexibility is most easily understood in a quantum optics context: linear interferometers~\footnote{The merging and splitting of system output fields can also refer to frequency conversion, so that the `linear interferometer' is to be understood in the most general sense.}, described by ${\bf S}$, take the field outputs of the system as inputs while $\pmb{\beta}$ represents the addition of weak local oscillators (WLOs) to the interferometer outputs prior to photodetection.  
We are interested in the case where the LOs are 
weak (that is, $|{\beta}_l|^2$ not much larger than $\Tr[c\dg_l c_l \rho_{\rm ss}]$) \blk so that a non-negligible
amount of information is gathered concerning the system upon each detection.  This means that the quantum trajectory is jump-like rather than diffusive; consequently, the PRE 
can consist of a finite number of states.    
We stress that most jump-like unravellings will lead, like diffusive unravellings, to PREs of infinite size. Those that lead to finite PREs, as we are interested in, are exceptional.

It was shown in~\crf{wiseman2001inequivalence} that an ensemble $\{ \wp_k,\, \ket{\phi_k} \}$ of size $K$ is physically realizable iff there exist real valued transition rates $\kappa_{jk} \geq  0$ (which naturally determine the occupation probabilities $\wp_k$) such that 
\beq
\label{jumpCond} 
  \forall k, \ {\cal L}\ket{\phi_k}\bra{\phi_k} = \sum_{j=1}^K \kappa_{jk} 
 \left(\ket{\phi_j} \bra{\phi_j}-\ket{\phi_k} \bra{\phi_k} \right).
\eeq
For each $k\in{1...K}$, \erf{jumpCond} is a matrix of constraints.  In general, a ME will allow multiple solutions to \erf{jumpCond} via the experimental freedom described by \erfs{jumpOps}{noJumpOps}; see below for further discussion. Most of the difficulty of our research program arises due to the system of non-linear constraints defined by \erf{jumpCond} being difficult to solve, even numerically, when $D>2$.

The transition rates $\kappa_{jk}$ play the role of free parameters in \erf{jumpCond}, and their possible values are determined by the range of solutions that exist.  However, we can make a few observations, without directly solving \erf{jumpCond}, based on simple requirements of the PRE.  Firstly, it is feasible that some of the $\kappa_{jk}$ can be zero only if this still allows the entire ensemble to be explored repeatedly in the long time limit.  That is, by following the possible non-zero transitions, all ensemble members can be reached from every other ensemble member.  A particularly simple arrangement of transitions that satisfies this is of a {\it cyclic} nature: the only non-zero transitions are of the form $\kappa_{{\rm mod}(k+1,K),k}$.  This is not to say that cyclic PREs will always exist, merely that they do satisfy the principle just discussed.  If further non-zero transition rates exist (or the pattern of transition rates is not based on a cycle), then we will refer to the PRE as {\it non-cyclic}.  Many of the example PREs that we analyze will be of a cyclic nature.

Given a viable PRE [an ensemble $\{ \kappa_{jk}\geq 0,\, \ket{\phi_k} \}$ satisfying \erf{jumpCond}],
there must exist an appropriately applied measurement scheme
such that the conditioned
state of the quantum system will, in the long-time limit, \blk jump between the ensemble members, spending a time in the state $\ket{\phi_k}$ proportional to $\wp_k$.  The measurement scheme in question will, in general, be adaptive, meaning that \blk the experimental setting parameters ($\pmb{\beta}$, ${\bf S}$), must be changed according to which state 
($k$) the system is currently in.
Adaptive measurements by a controllable local oscillator were first studied in the context of quantum jumps in Ref.~\cite{wiseman1999quantum}. They are more commonly studied for their use in state discrimination~\cite{Dol73} 
and phase estimation~\cite{PhysRevLett.75.4587,BerHalWis13}, 
with the latter having been realized experimentally  ~\cite{PhysRevLett.89.133602,PhysRevLett.104.093601,YH&12}.

In the body of our paper, we focus on characterizing PREs for MEs possessing a symmetry, and defer to the appendices an analysis of the measurement schemes required to produce the PREs.  The reason for this is that finding the measurement scheme for a given PRE is a comparatively simple computational task, whereas we are concerned with ameliorating the difficulty of finding PREs in the first place.

\subsection{Generalized Bloch representation}

The generalized Bloch representation of $N$-level quantum states will prove very useful when discussing the symmetries of PREs.  In $D=2$, it is used also in describing analytical PREs and as a visualization tool.  In this subsection, we introduce the Bloch vector for $N$-level systems, with these purposes in mind.  Our first step is to lay out some notation.

\blk
Let $\mathfrak{B}\left({ \mathbb{H}}\right)$ represent the set of bounded linear \blk operators on the Hilbert space $\mathbb{H}$.  Density matrices, $\rho$, being trace-one positive-semidefinite operators, form a convex set 
$\mathfrak{D}\left({ \mathbb{H}}\right)\subset\mathfrak{B \blk}\left({\mathbb{H}}\right)$, with extreme points in the set corresponding to one-dimensional projectors, 
$\ket{\psi}\bra{\psi}$, that represent pure states. 
An operator basis for $\rho$, say $\hat{{\pmb \sigma}}=\{\hat{\sigma}_{j}\}_{j=1}^{D^2}$, can be chosen that consists of $D^2$ 
 orthogonal Hermitian matrices, $D^2-1$ of which are traceless.  The generalized Pauli matrices, suitably normalized, provide an example
of one such basis~\cite{KIMURA2003339}.  In the chosen basis, the state of the system can then be represented by $D^2$ coordinates, comprising a real-valued vector, $ \pmb{r}=\{r_{j}\}_{j=1}^{D^2}\in \mathbb{R}^{D^2}$, as follows: 
\beq
\rho= \frac{1}{D} \sum_{j=1}^{D^2}r_{j}\hat{\sigma}_{j}.
\label{genBloch}
\eeq

Commensurately, the Lindblad dynamics of \erf{me1} can be stated as
\beq
\dot{ \pmb{r}}={\bf L}\pmb{r},
\label{lindVec}
\eeq
with ${\bf L}$ being a real, $D^2\times D^2$ matrix representation of the Lindbladian superoperator, ${\cal L}$ (${\bf L}$ should not be confused with the number of decoherence channels, $L$).  We can simplify further by setting the traceful operator of our basis to be
$\hat{\sigma}_{D^2}=\mathbb{1}$, as then $\dot{r}_{D^2}=0$ and $r_{D^2}=1$.  The dynamics of \erf{lindVec} reduces to the $D^2-1$ dimensional subspace ($\pmb{x} =\{r_{j}\}_{j=1}^{D^{2}-1}$, corresponding to the traceless operator basis components), giving
\beq
\dot{ \pmb{x}}={\bf L}_{0} \pmb{x}+ \pmb{b},
\label{lindVecReduced}
\eeq
where ${\bf L}_{0}$ is the restriction of the Lindbladian matrix representation, ${\bf L}$, to the first $D^2-1$ coordinates and $ \pmb{b}$ is its final column (that is $b_{j}={\bf L}_{j,D^2}$ for $j<D^2$)~\footnote{The expressions for ${\bf L}_{0}$ and \pmb{b} in terms of $\hat{H}_{\rm eff},\hat{c}_{l}$ in the generalized Pauli basis are given in Ref.~\cite{PhysRevA.81.062306}, although note that our ${\bf L}$ includes the irreversible as well as Hamiltonian evolution and that we have used a different normalization.}.
The $D^{2}-1$ dimensional vectors $\pmb{x}$, which are called generalized Bloch vectors~\cite{1751-8121-41-23-235303}, occupy the region of $\mathbb{R}^{D^2-1}$ that is mapped to $\mathfrak{D}\left({ \mathbb{H}}\right)$ via \erf{genBloch} with  $r_{D^2}=1$. This region is contained within a closed $(D^{2}-1)$-ball of radius $\sqrt{\blk D(D-1)/2}$. 
(All geometric `balls'  --- and `discs' --- in our work should be understood as `closed'.) 

In $D=2$, we have the familiar Bloch ball, every point of which is a valid density matrix.  However, for $D>2$, the inverse mapping of $\mathfrak{D}\left({ \mathbb{H}}\right)$, to the vectors $\pmb{x}$, gives only a convex compact subregion of the ball (that is still of dimension $D^2-1$).  Moreover, for $D>2$, the boundary of this convex subregion will comprise both extremal (pure) and non-extremal (mixed) states.  Extremal states are those that cannot be formed from a combination (mixture) of other states in the region; a collection of such states is called an extremal set. 

Many of the constraints defining PREs can be expressed in the Bloch representation. The assumption that ${\cal L}$ has a unique steady state, $\rho_{{\rm ss}}$, of rank $D$, is equivalent to $\pmb{x}_{{\rm ss}}=-{\bf L}_{0}^{-1}\pmb{b}$ being unique.  Then, using \erf{lindVecReduced}, we obtain the Bloch representation of \erf{jumpCond}
\beq
 \forall k,\,\,\,\,{\bf L}_{0} \pmb{x}_{k}+ \pmb{b}=
\sum_{j=1}^K \kappa_{jk} 
 \left(\pmb{x}_{j}-\pmb{x}_{k} \right),
\label{jumpCondVec}
\eeq
for states on the surface of the generalized Bloch ball:
\beq
 \forall k,\,\,\,\, \pmb{x}_{k}.\pmb{x}_{k}=\sqrt{D(D-1)/2}.
\label{jumpCondVec2}
\eeq
As discussed above, for $D>2$, these are only necessary conditions for $\left\{\kappa_{jk},\pmb{x}_{k}\right\}$ to represent a PRE; additionally $\left\{\pmb{x}_{k}\right\}$ must lie on the portion of the Bloch sphere that corresponds to valid density matrices.

\subsection{Cyclic qubit PREs}
Throughout this paper, extensive use will be made of $D=K=2$ PREs (which are necessarily cyclic) to illustrate the developed symmetry tools.  This subsection will serve the dual purpose of pedagogically introducing the reader with less background knowledge to some actual PREs and, also, provide an example that will be frequently referenced in later sections. 

Some analytical results, discovered in \cite{karasik2011many,karasik2011tracking}, exist concerning $K=2$, $L=1$ qubit PREs.  
The evolution consists of periods of smooth dynamics (during which there are no detections), under the action of the $\hat{H}^{\prime}_{\rm eff}$, interspersed with jumps caused by detection events, described by the single Lindblad operator $\hat{c}^{\prime}$ (see \erfs{jumpOps}{noJumpOps}).  Continuing our exposition within a quantum optics context, the experimental freedom is parameterized by the amplitude, $\beta$, of a single WLO; the value of $\beta$ will be switched dependent upon which PRE state is occupied. Taking the system to be in state, $\ket{\phi_1}$, the no-jump and jump operators are given by 
\bqa
\hat{H}^{\prime}_{\rm eff}(\beta_{1})&=&\hat{H}-\frac{i}{2}\hat{c}^{\dag}\hat{c}-i\beta_{1}^{*}\hat{c}-i\frac{|\beta_{1}|^{2}}{2}\\
\hat{c}^{\prime}(\beta_{1})&=&\hat{c}+\beta_{1}.
\eqa
Those for $\ket{\phi_2}$ are obtained merely by replacing $\beta_{1}$ with $\beta_{2}$.  The $K=2$ PRE is then defined by 
\bqa
\hat{H}^{\prime}_{\rm eff}(\beta_{k})\ket{\phi_k}\propto \ket{\phi_k}\\
\hat{c}^{\prime}(\beta_{k})\ket{\phi_k}\propto\ket{\phi_{{\rm mod}(k,2)+1}}\label{cyclic2}.
\eqa
In other words, the state $\ket{\phi_k}$ is an eigenstate of the no-jump operator $\hat{H}^{\prime}_{\rm eff}(\beta_{k})$ and the jump operator, $\hat{c}^{\prime}(\beta_{k})$ takes the state $k$ to the state ${\rm mod}(k,2)+1$.  Of course, after two jumps, the state must return to its initial state, which implies that
\beq
\hat{c}^{\prime}(\beta_{2})\hat{c}^{\prime}(\beta_{1})\ket{\phi_1}=
\left[(\beta_{1}\beta_{2}+\left(\beta_{1}+\beta_{2} \right) \hat{c}+\hat{c}^{2}\right]\ket{\phi_1}
\propto\ket{\phi_1}.\label{doubleJump}
\eeq
Taking $\hat{c}$ to be traceless (without loss of generality) implies that $\hat{c}^{2}\propto\mathbb{1}$.  Consequently, provided we exclude the case where $\ket{\phi_1}=\ket{\phi_2}$, \erf{doubleJump} can only be fulfilled if $\beta_{1}=-\beta_{2}$.  The actual value of $\beta_{1}$ can be determined in terms of the states $\ket{\phi_k}$.  The $\ket{\phi_k}$ themselves can be analytically found, as we now indicate.

For $K=2$, \erf{jumpCondVec} reduces to the single eigenvalue equation (for arbitrary $L\geq 1$)
\beq
{\bf L}_{0}( \pmb{x}_{1}-\pmb{x}_{2})=
\left(\kappa_{12}+\kappa_{21}\right) ( \pmb{x}_{2}-\pmb{x}_{1})
\eeq
and we can conclude that the two PRE states are given by
\bqa
\pmb{x}_{1}=\pmb{x}_{{\rm ss}}+\eta_{1} \pmb{e}\label{analyticPRE1}\\
\pmb{x}_{2}=\pmb{x}_{{\rm ss}}-\eta_{2} \pmb{e},
\label{analyticPRE2}
\eqa
where $\pmb{e}$ is a real eigenvector of ${\bf L}_{0}$ and $\eta_{1,2}$ are scalars dependent upon $ \pmb{x}_{{\rm ss}}$ and $\pmb{e}$ (see \cite{karasik2011many} for the expressions).  For $K>2$, there is no general analytic form of the PRE and numerical methods are used. 


\subsubsection{Resonance fluorescence}
\label{RFa}
 ${}$\blk To give a specific example, we take the resonance fluorescence ME that was analyzed, in detail, in Refs.~\cite{karasik2011many,karasik2011tracking}.  In this system, the fluorescence of a resonantly driven two-level system is coherently mixed with a single WLO, at the resonance frequency, before detection. The ME is of the form \erf{me1}, with $\hat{H}=\frac{\Omega}{2}\hat{\sigma}_{x}$ and a single Lindblad jump operator $\hat{c}=\sqrt{\gamma}\ketbra{0}{1}$.  For later convenience, we perform a change of basis relative to \cite{karasik2011many,karasik2011tracking}, such that $\ket{0}\rightarrow i\ket{0}$.  The rebit plane then becomes the $y$-$z$ plane of the Bloch ball.  In other words, $\hat{\sigma}_{x}$ has been chosen as the imaginary Pauli operator, so that $\hat{H}=\frac{\Omega}{2}\hat{\sigma}_{x}=i\Omega\left(\ketbra{0}{1}-\ketbra{1}{0}\right)/2$\blk .

Working in the Pauli basis, $\{\sigma_{x},\sigma_{y},\sigma_{z}\}$, we find that the expressions for ${\bf L}_{0}$ and $\pmb{b}$ (appearing in \erf{lindVecReduced}) are \blk
\beq
{\bf L}_{0}=\left[
\begin{array}{ccc}
   -\gamma/2       &0&0   \\
    0     			 & -\gamma/2&-\Omega\\
    0				&\Omega	     &-\gamma
\end{array}\right],
\quad \pmb{b}=\left[
\begin{array}{c}
   0         \\
    0        \\
   - \gamma				
\end{array}\right],
\label{LBlockResFlu}
\eeq
from which we can determine the steady state $\pmb{x}_{{\rm ss}}=(0,2\gamma\Omega,-\gamma^2)^{{\rm T}}/(\gamma^2+2\Omega^2)$.  
The three \blk right-eigenvectors of ${\bf L}_{0}$ are: $ \pmb{e}_1=(1,0,0)^{{\rm T}}$
and $ \pmb{e}_{\pm}=(0,\gamma\pm\sqrt{\gamma^2-16\Omega^2},4\Omega)^{{\rm T}}$ 
(the latter being unnormalized).
When all eigenvectors are real ($\gamma^2-16\Omega^2\geq 0$), each eigenvector provides a $K=2$ PRE, specified by \erfs{analyticPRE1}{analyticPRE2}.  We display each of the $K=D=2$ PREs in
\frf{2KPREs}(a)-(c), with subplots (a),(b) associated with $ \pmb{e}_{\pm}$ and subplot (c) with $ \pmb{e}_{1}$~\footnote{\label{footRef2}\frf{2KPREs}(a)-(c) \blk appear in Ref.~\cite{karasik2011many} (with slightly different parameters), but are reproduced to make our paper more self-contained.  Numerical calculations were independently carried out (for verification purposes) using MAGMA computational algebra software~\cite{bosma1997magma} and figures were created using QUTIP~\cite{JOHANSSON20131234}.}. 


\begin{figure}[t]
\centering
 \includegraphics[width=0.6\textwidth]{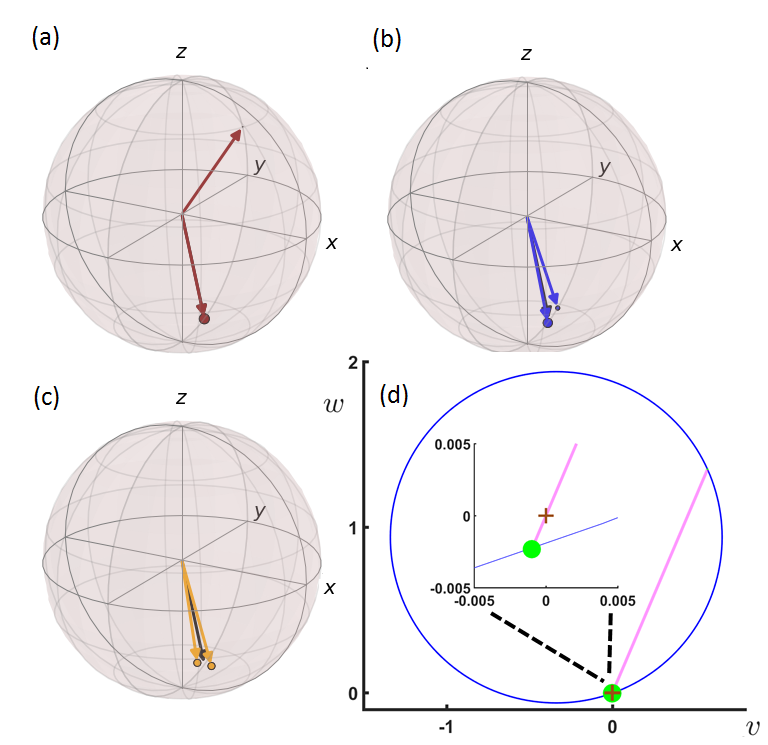}
\caption{Coloured arrows show Bloch vectors for all three $K=2$ PREs (one in each subplot (a-c)) of the ME of \erf{LBlockResFlu} with $\Omega/\gamma=0.18$.  The volume of the ball at the tip of each PRE state arrow represents the probability $\wp_k$ that the qubit occupies the corresponding pure state. 
The steady state, $\pmb{x}_{{\rm ss}}$, is plotted as a black arrow but it is not often visible as it lies very close to the overlaid PRE states --- for example, in subplot (a) it lies almost coincident with the highly occupied PRE member. In subplot (d) (which relates to material introduced in \srf{invSubspaceSec}), the PRE of subplot (a) is reproduced.  As it lies entirely in the $u=0$ subspace of $\mathfrak{U}_0$, only this subspace is shown, the boundary of which is the blue circle, upon which lie pure states.  \blk
In the space $\mathfrak{U}_0$, the steady state lies at the origin, marked by a brown +. The green discs show the PRE state locations, with their area proportional to their occupation probability.  The brown + is almost concentric with the larger green disc, so we also include the figure inset, which zooms in on the region of the origin.  Note that we have not magnified the area of the green disc, in order to highlight the separation of the PRE state from the brown cross (when viewed closely).   The steady state is only marginally mixed and consequently lies very close to the circle of pure states. The magenta secant marks $\mathfrak{I}_0$ as a line through the origin meeting the surface at two pure states. 
}
\label{2KPREs}
\vspace{-5mm}

\end{figure}

%

\blk

\section{Feasibility of finding PREs}
\label{feas}
As this paper's purpose is to make easier the solution of \erf{jumpCond}, it is important for us to motivate this as being necessary.  Further details will be provided in \srf{utility}, but let us here consider the difficulty of finding PREs in $D=3$, which is the smallest dimension for which it is not possible, in general, to construct an example PRE via analytic means.  For $D=3$, it is argued, in Ref.~\cite{karasik2011many}, that the minimum expected PRE size is $K=5$.  Thus, \erf{jumpCond} presents as $5$ coupled matrix equations, each of which represents $D^{2}=9$ constraints, giving $45$ total constraints.  The RHS of \erf{jumpCond} consists of terms cubic in the variables describing the transition rates (linear contribution) and the pure state vectors (quadratic contribution).  

How difficult is it to solve a system of $45$ coupled, mostly cubic, polynomials?  The answer is that it is very difficult, at least as judged by current computational resources.   
In~\crf{courtois2000efficient} it is suggested that even when using the highly efficient Gr{\"o}bner basis Faug{\`e}re F4 algorithm~\cite{faugere1999new} (a variant on the Buchberger algorithm~\cite{buchberger1976theoretical}) quadratic systems of size larger than 15 equations are very difficult.  The constraints of \erf{jumpCond} are cubic (except for $K$ quadratic normalisation constraints) and are, resultantly, even more difficult.  As another example, the computational algebra software MAGMA~\cite{bosma1997magma} has timings~\cite{MAGMAtiming} provided for several benchmark problems, with the Cyclic 9 problem (see~\cite{faugere2001finding} and references therein) over a rational field listed as being solved in 4.6 days.  The Cyclic 9 problem has only 9 equations but has a maximum degree of 9.  Our experience with Gr{\"o}bner basis techniques using MAGMA was that an example system of 16 equations was soluble in about 13hrs~\cite{privateSteel} whereas size 20 systems were out of reach (limitations of 200Gb of Ram were exceeded after several days runtime).  

In the current paper, we defer the numerical task of solving \erf{jumpCond} for $D>2$ and, instead, after acknowledging its difficulty, focus on developing techniques that will allow such problems to be tackled in future work (where we will also provide more details on the available numerical methods).  To test our new methods, we consider in detail mostly $D=2$ examples and view these from the new perspective provided by an analysis of the available symmetry.

\section{Invariant subspaces of ${\cal L}$}
\label{invSubspaceSec}

\subsection{Definitions}
\label{defs}

In this section, we will define what is meant by an invariant subspace of ${\cal L}$.  To do so, 
a co-ordinate translation is made.  Recall that we assume ${\cal L}$ to have a unique steady state, $\rho_{{\rm ss}}$, of rank $D$, 
\beq
\rho_{{\rm ss}}=\frac{1}{D}\mathbb{1}+\frac{1}{D} \sum_{j=1}^{D^2-1}(x_{{\rm ss}})_{j}\hat{\sigma}_{j}.
\label{rhoSS}
\eeq
where 
\beq
\pmb{x}_{{\rm ss}}=-{\bf L}_{0}^{-1}\pmb{b}.
\label{steadyState}
\eeq
Now let us translate to coordinates $ \pmb{u}=\pmb{x}-\pmb{x}_{{\rm ss}}$, so that 
\beq
\rho=\rho_{{\rm ss}}+\frac{1}{D}\sum_{j=1}^{D^2-1}u_{j}\hat{\sigma}_{j}.
\label{rhoReduced}
\eeq
For these new coordinates, \erf{lindVecReduced} becomes
\beq
\dot{ \pmb{u}}={\bf L}_{0} \pmb{u}
\eeq
and the new steady-state, $\pmb{u}_{{\rm ss}}$, is at the origin.

As the vectors $\pmb{u}$ are merely translations of $\pmb{x}$, we will talk about them in the same geometric terms as the generalized Bloch vectors.  The region of $\mathbb{R}^{D^2-1}$ that they occupy will be referred to as $\mathfrak{U}_{0}$, which is, in turn, mapped to $\mathfrak{D}\left({ \mathbb{H}}\right)$ via \erf{rhoReduced}.  The boundary of the space $\mathfrak{U}_{0}$ (as viewed in the new coordinate system) will lie within (or on) that of a sphere of radius $\sqrt{\blk D(D-1)/2}$ translated by $-\pmb{x}_{{\rm ss}}$ from the origin.  Many of the example systems in this paper will have $D=2$; in such cases, to make it clear which coordinate system is being discussed, we will use $x,y,z$ as the components of $\pmb{x}$ (which lies in the Bloch ball) and $u,v,w$ as the components of $\pmb{u}$ (which lies in $\mathfrak{U}_{0}$). 

We define an ${\cal L}$-invariant subspace $\mathfrak{D}_\mathfrak{I}  \subset \blk  \mathfrak{D}\left({ \mathbb{H}}\right)$ to be a convex space containing $\rho_{\rm ss}$, and having the property that the image of $\mathfrak{D}_\mathfrak{I}$ under $e^{{\cal L}t}$ for any $t\geq 0$ is $\subseteq \mathfrak{D}_\mathfrak{I}$~\footnote{It is important to realize that our notion of invariance is different from that of Ref.~\cite{4639467}.  We consider invariant subspaces of $\mathfrak{D}\left({ \mathbb{H}}\right)$, 
with dim$(\mathbb{H})=D$.
We require that the invariant space contains rank $D$ density matrices (meaning that there will exist density matrices that are formed by an ensemble of $D$ linearly independent pure state projectors).
This is in contrast to Ref.~\cite{4639467}, where the invariant subspace, of Hilbert space dimension $\tilde{D}$, is $\mathfrak{D}( \tilde{\mathbb{H}})$  for $\mathbb{H}= \tilde{\mathbb{H}}\oplus
{\mathbb{H}_{R}}
$ (with, here, $R$ being the remainder space).  In other words, their invariant space consists of all possible density matrices with support solely in some Hilbert subsystem, $\tilde{\mathbb{H}}$, whereas that considered in this paper is a compact subregion $\mathfrak{D}_{\mathfrak{I}}\subset\mathfrak{D}\left( \mathbb{H}\right)$.
Note that in the long time limit, under ME dynamics, the state necessarily relaxes to the unique $\rho_{{\rm ss}}$, a single point within the invariant subspace.}.  
Equivalently, this ${\cal L}$-invariant subspace can be represented by a convex subregion $\mathfrak{I}_{0}\subset\mathfrak{U}_{0}$, comprising vectors of the form $\pmb{u}$, such that the image of $\mathfrak{I}_{0}$ under $e^{{\bf L}_{0}t}$ is contained within $\mathfrak{I}_{0}$ for $t\geq 0$.
That is,  $\mathfrak{I}_{0}$ characterizes the invariant subspace and we will term the dimension of $\mathfrak{I}_{0}$ as the invariant subspace dimension.

For the invariant subspace definition to be an interesting one in the context of PREs,
we require that $\mathfrak{D}_{\mathfrak{I}}$ corresponds to an $\mathfrak{I}_{0}$ that is an $N$-dimensional projective subspace of $\mathfrak{U}_{0}$, with $D-1\leq N < D^2-1$. The lower bound on dimensionality is necessary so that it is 
possible for $\mathfrak{I}_{0}$ to represent a PRE (which must contain $D$ linearly independent pure states).  The upper bound will allow a simplification in our search for PREs, due to the reduced dimensionality.  That $\mathfrak{I}_{0}$ is a projective subspace indicates that it consists of lines through the origin; this ensures that the steady state is included (the origin in $\mathfrak{U}_{0}$ space) and that the extremal set of $\mathfrak{I}_{0}$ is in the boundary of $\mathfrak{U}_{0}$.  However, as discussed earlier, the boundary of $\mathfrak{U}_{0}$ is not necessarily pure, so we additionally require that the extremal set of $\mathfrak{I}_0$ is in the extremal set of $\mathfrak{U}_0$. 

To span the entire density matrix space, which is represented by $\mathfrak{U}_{0}$, requires
the complement to $\mathfrak{I}_0$, an orthogonal projective subspace of $\mathfrak{U}_{0}$ (we also include the origin) that is of dimension at most $D^2 - D$.  We term this space $\mathfrak{R}_{0}$ and note that $\mathfrak{I}_{0}\cup\mathfrak{R}_{0}=\mathfrak{U}_{0}$ and $\mathfrak{I}_{0}\cap\mathfrak{R}_{0}={\bf 0}$ (the origin, not the empty set, $\emptyset$).  Let us now move to a new orthonormal basis for $\mathfrak{U}_{0}$ such that each basis vector lies wholly within either $\mathfrak{I}_{0}$ or $\mathfrak{R}_{0}$.  Additionally, we write the coordinates of $\pmb{u}$  with those $N$ degrees of freedom corresponding to $\mathfrak{I}_{0}$ first, followed by those of $\mathfrak{R}_{0}$.
The purpose of this is to clearly expose the invariant subspace 
in the representation ${\bf L}_{0}$, which will take on the block form
\beq
{\bf L}_{0}=\left[
\begin{array}{c|c}
    {\bf L}_{\mathfrak{I}_{0}}       & {\bf L}_{{\mathfrak{I}_{0},\mathfrak{R}_{0}}}\blk   \\
\hline
    {\bf 0}\blk     			 & {\bf L}_{\mathfrak{R}_{0}  }
\end{array}\right],
\label{LBlock}
\eeq
where ${\bf L}_{\mathfrak{I}_{0}}$ and ${\bf L}_{\mathfrak{R}_{0}}$ are square with length
$N$ and $(D^2-N-1)$ respectively and ${\bf L}_{{\mathfrak{I}_{0},\mathfrak{R}_{0}}}$ is $N\times (D^2-N-1)$.  It is clear, from \erf{LBlock}, 
that $\pmb{u}$ will not be mapped outside $\mathfrak{I}_{0}$ as long as it is initialized inside it. The upper-right block, ${\bf L}_{{\mathfrak{I}_{0},\mathfrak{R}_{0}}}$, of the representation ${\bf L}_{0}$ in \erf{LBlock} is non-zero in general, reflecting that $\mathfrak{R}_{0}$ may not itself be invariant.

To complete this subsection, a procedure for finding an invariant subspace is described.
Firstly, we choose a basis for the $D^2-1$ traceless, orthogonal, Hermitian operators $\hat{{\pmb \sigma}}=\{\hat{\sigma}_{j}\}_{j=1}^{D^2-1}$ (we have already set $\hat{\sigma}_{D^2}=\mathbb{1}$).  This directly leads to expressions for the matrix representations ${\bf L}$ and ${\bf L}_{0}$ (see, for example, Ref.~\cite{PhysRevA.81.062306}).  We now examine ${\bf L}_{0}$ and calculate its right-eigenvectors.  The case for which ${\bf L}_{0}$ does not have $D^2-1$ linearly independent right-eigenvectors (that is, ${\bf L}_{0}$ is not diagonalizable~\cite{anton2010elementary}) is discussed in \cref{notDiagonalizable}.  These eigenvectors will define directions in $\mathfrak{I}_{0}\subset\mathfrak{U}_{0}$.  In general, the right-eigenvectors will be complex-valued, and appear as complex conjugate pairs, but we can form a linearly independent real-valued pair of vectors by taking their real and imaginary parts respectively.  
Each of these real-valued pair of vectors defines a plane, that will form an invariant subspace.  Additionally, each real eigenvalue of ${\bf L}_{0}$ will define an invariant subspace via its corresponding real eigenvector.  
Obviously we can then form larger invariant spaces by combining any of the smaller invariant spaces --- this will be necessary, in general, as we require $\mathfrak{I}_{0}$ to be of dimension at least $D-1$.  The eigenvectors of ${\bf L}_{0}$ will not typically be orthogonal, so that they will not provide an orthogonal basis for $\mathfrak{U}_{0}$.  Consequently, the final step is to find such an orthogonal basis, 
with each basis vector belonging solely either to $\mathfrak{I}_{0}$ or $\mathfrak{R}_{0}$.  
As mentioned, there will typically be eigenvectors of ${\bf L}_{0}$ that are linearly independent from, but not orthogonal to, $\mathfrak{I}_{0}$ --- these will not then span $\mathfrak{R}_{0}$, meaning that $\mathfrak{R}_{0}$ is not itself an invariant subspace.  In the 
final discussed basis, ${\bf L}_{0}$ will take on the form given in \erf{LBlock}.

\subsection{Utility}
\label{utility}
Once an appropriate invariant subspace has been identified, it is natural to look for solutions to \erf{jumpCond} that lie entirely in this subspace.  That is, 
\beq
  \forall k,\quad \ket{\phi_k}\bra{\phi_k} \in\mathfrak{D}_\mathfrak{I},
\label{invSubspace}
\eeq
where, as a reminder, $\{\ket{\phi_k}\}_{k=1}^{K}$ are the PRE members (if the PRE exists). As $\ket{\phi_k}\bra{\phi_k}$ are pure states, they are extremal points of $\mathfrak{D}_\mathfrak{I}$.  
In terms of the space $\mathfrak{U}_{0}$, \erf{invSubspace} is equivalent to
\beq
  \forall k,\quad \pmb{u}_{k} \in\mathfrak{I}_{0},
\label{invSubspaceFrak}
\eeq
where $\{\pmb{u}_{k}\}_{k=1}^{K}$ describe the PRE members.  Note that the requirement that $\{\pmb{u}\}_{k=1}^{K}$ correspond to pure states is enforced for $D=2$ via the constraint $||\pmb{x}_{k}||^{2}=1=||\pmb{u}_{k}+\pmb{x}_{{\rm ss}}||^{2}$, $\forall k$.  For $D>2$ further constraints are imposed~\cite{KIMURA2003339}.

Assuming  that we have made a change of basis such that ${\bf L}_{0}$ is of the form \erf{LBlock}, the form of \erf{jumpCond} in $\mathfrak{U}_{0}$ space is 
\beq
 \forall k,\,\,\,\,
\left[
\begin{array}{c|c}
  {\bf L}_{\mathfrak{I}_{0}}       & {\bf L}_{{\mathfrak{I}_{0},\mathfrak{R}_{0}}}\blk   \\
\hline
      {\bf 0}\blk    			 & {\bf L}_{\mathfrak{R}_{0}  }
\end{array}\right]
\pmb{u}_{k}=
\sum_{j=1}^K \kappa_{jk} 
 \left(\pmb{u}_{j}-\pmb{u}_{k} \right).
\label{blockGOE2}
\eeq
The utility of assuming $\pmb{u}_{k} \in\mathfrak{I}_{0}$ is now made clear as, since $\pmb{u}_{k}$ has only non-zero components in its first $N={\rm dim}(\mathfrak{I}_{0})$ coordinates, \erf{blockGOE2} simplifies to 
\beq
 \forall k,\,\,\,\,
{\bf L}_{\mathfrak{I}_{0}}\pmb{u}_{k}|_{\mathfrak{I}_{0}}=
\sum_{j=1}^K \kappa_{jk} 
 \left(\pmb{u}_{j}|_{\mathfrak{I}_{0}}-\pmb{u}_{k}|_{\mathfrak{I}_{0}} \right),
\label{blockGOE2simp}
\eeq
with $\pmb{u}_{k}|_{\mathfrak{I}_{0}}$ being the restriction of $\pmb{u}_{k}$ to the domain of $\mathfrak{I}_{0}$.
\blk
In other words, the constraint equations of \erf{jumpCond} relating to $\mathfrak{R}_{0}$ 
are trivially satisfied, 
thus reducing the size of the polynomial system that defines the PRE.  Specifically, the number of constraint equations (including normalization constraint) is reduced from $KD^{2}$ to $K(N+1)$ with $N<D^{2}-1$.  This should be considered in the light of the exponential (or worse) scaling of the computational difficulty of solving polynomial systems in the system size.  Another perspective is given by the fact that the B{\'e}zout bound (for the maximum number of solutions to a polynomial system) is given by the product of the largest degree of the polynomial equations.  The B{\'e}zout bound is therefore a factor of $3^{D^{2}-1-N}$ smaller for the polynomial system defining the existence of PREs lying wholly within the invariant subspace.  Additionally, the polynomials are more sparse than if the full space were being considered.  Both of these considerations reduce the computational cost of finding solutions to the constraints.  

There are, in general, multiple solutions to \erf{jumpCond} for a given $K$.  Some fraction of these (which may be zero, one or in between) will lie entirely within $\mathfrak{D}_\mathfrak{I}$ 
 --- it is these solutions that we are focusing on here. For $N$ not too large it should be possible to do an exhaustive search to find all of them, if they exist, or prove that there are none.  Whether any are found or not says nothing about whether solutions {\it not} belonging to $\mathfrak{D}_\mathfrak{I}$ exist.  That is, solutions $\{\ket{\phi_k}\bra{\phi_k}\}$ having any non-zero operator support outside of $\mathfrak{D}_\mathfrak{I}$ will not be found with this approach, a point that will be illustrated later in our work.

An important consideration is whether it is to be {\it expected} that solutions of \erf{jumpCond} satisfying \erf{invSubspace} can be found --- this consideration being distinct from the fact that they will be easier to find if they do exist.  That is, is the dimension of $\mathfrak{D}_\mathfrak{I}$ sufficient to provide enough free variables to satisfy all the constraint equations of \erf{jumpCond}?  In Ref.~\cite{karasik2011many}, where invariant subspace considerations were {\it not} made, it was argued heuristically, via the counting of free parameters and constraints of \erf{jumpCond}, that typically 
\beq 
K\geq D^2-2D+2
\label{heuristic1}
\eeq
 ensemble states are required for a PRE to be possible.  This was based on $D^2-1$ constraints per ensemble member, $K^2-K$ transition rates and $2D-1$ unknowns to describe each ensemble state.  The restriction of $\{\ket{\phi_k}\bra{\phi_k}\}$ to some subset of the extremal set of $\mathfrak{D}\left({ \mathbb{H}}\right)$, as appropriate when considering an invariant subspace, can be achieved by placing constraints on the $2D-1$ state variables.  Because of the quadratic dependence of the number of constraints upon the size of $\mathfrak{D}_\mathfrak{I}$ it is possible to reduce their number faster (in the case of a qubit, at an equal rate) than the those of the free parameters when an invariant subspace (in the sense defined above) is considered.  In other words, by reducing the scope of the PRE search (to that of an invariant subspace) it becomes apparent that it can be easier to satisfy a parameter and constraint counting heuristic for PRE existence.  What may have been an intractable task (due to the computational complexity of finding PREs) is made possible, with the cost being that some PREs (those, if any, that are not contained in $\mathfrak{D}_\mathfrak{I}$) are not discoverable in this way.  An example of the simplification provided by invariant subspaces, for the finding of PREs, is provided in the next section.
 
In general, a given ME will possess a number of invariant subspaces.  How then do we choose which ones to investigate for the presence of PREs?  An initial criterion is that they are `interesting' in the sense already discussed.  That is, we should consider only $N$-dimensional projective subspaces of $\mathfrak{U}_{0}$, with $D-1\leq N < D^2-1$.  Furthermore, we would limit to invariant subspaces for which PREs can be expected to be found, as per the previous paragraph in terms of parameter and constraint counting.  This process of elimination would depend on the desired ensemble size, $K$, as this affects the counting heuristics.  Often we will desire the simplest computational search for PREs; this mandates the choice of minimally sized (described by the fewest parameters) invariant subspaces as being the initial search targets.  
Note that the size of $L$ in comparison to $D$ may affect the minimum feasible $K$ --- this will be investigated in a later publication~\cite{warWis}.

\blk

\subsection{An important class, and its constraint and parameter count}
\label{realRho}
An important invariant subspace is the conceptually (and computationally) simple one of real-valued density matrices.  That is, some predetermined basis exists in which $\mathfrak{D}_\mathfrak{I}=\R^{D\times D}\cap\mathfrak{D}\left({ \mathbb{H}}\right)$.  
This subspace, of dimension $N=(D^{2}+D)/2-1$, \blk provides an example of the minimum sized invariant subspace that can be formed that is consistent with \erf{invSubspace} without constraining the Hilbert space dimension of span($\{\ket{\phi_k}\}$) to be less than $D$. 
This is because the extremal subset of $\mathfrak{D}_\mathfrak{I}=\R^{D\times D}\cap\mathfrak{D}\left({ \mathbb{H}}\right)$ still contains $D$ orthogonal states --- pure `redit' states that are each representable as a ray in a $D$-dimensional real Hilbert space. A minimal invariant subspace (in the sense described above) is desired in order to reduce, to the greatest extent possible, the polynomial system represented by \erf{jumpCond}. Of course $\mathfrak{D}_\mathfrak{I}=\R^{D\times D}\cap\mathfrak{D}\left({ \mathbb{H}}\right)$ will only be immediately relevant when ${\cal L}$ possesses such an invariant subspace, but the concepts --- of choosing an as small as possible interesting invariant subspace and then parametrising this space --- are more general than this. 
Given $\mathfrak{D}_\mathfrak{I}=\R^{D\times D}\cap\mathfrak{D}\left({ \mathbb{H}}\right)$, we then search for an example PRE 
that is real valued in the utilized basis.  The imaginary constraints of \erf{jumpCond} are then automatically satisfied.  

Let us now give the new inequality (analogous to \erf{heuristic1}) that must be satisfied, for the number of parameters to equal or exceed the number of constraints, when a real-valued invariant subspace is assumed as well as a real-valued PRE. Considering a generic ME~\footnote{The effect of having a constraint on the number of Lindblad channels --- in particular of having $L< D-1$ --- will be explored in future work~\cite{warWis}.}, 
there are $K^2-K$ transition rates as before, but now only $D-1$ free parameters and $D^2-D$ constraints per ensemble state 
because we are restricting to real state-vectors and matrices. \blk Solving for the integer minimum ensemble size, $K$, such that the number of free parameters is at least as large as the number of constraints gives
\beq
K\geq \frac{1}{2}\left(D^{2}-D+2\right).
\label{heuristic2}
\eeq  
Comparing with \erf{heuristic1}, we see that the scaling with $D^2$ is still present, although the coefficient is reduced by a factor of two.  Hence, for large $D$, a search for PREs of roughly half the expected ensemble size indicated by the heuristic of Ref.~\cite{karasik2011many} may be justified.   In \trf{table1}, the minimal PRE size, $K$, is given, for small $D$, with and without the use of a redit invariant subspace.  Of course, for small $D$, the difference is less pronounced, but it may still be of great importance given the numerical difficulty of finding PREs. For $D=3$, we find, from \erf{heuristic2} (or \trf{table1}), that we can reasonably hope to find PREs with an ensemble size of $K=4$, compared with $K=5$ from \erf{heuristic1}.  It also follows that there are only 6 constraint equations (including normalisation constraint of the state) per ensemble member, giving a (square) polynomial system of size 24.  This is a vast improvement upon the size 45 system, that presents in the absence of an invariant subspace.  Additionally, the number of monomial terms is greatly reduced.  In the limit of large $D$, this reduction approaches a factor of $4$, as can be deduced by noting that the constraints have a quadratic dependence upon state-vector variables (expansion gives $4$ terms if complex-valued but only $1$ if real-valued).  Both simplifications give hope that future numerical investigations of PREs can extend to $D=3$ and beyond. The case of  $D=2$, describing a `rebit', actually leads to PREs of the same size as that predicted by \erf{heuristic1}.  However, even in the case of $D=2$, the analysis above leads us to a better understanding of the nature of the PREs; in our upcoming discussion we highlight the symmetry that exists in the $D=2$ PREs of Refs.~\cite{karasik2011many,1367-2630-16-6-063028}.   

Note the importance of the fact that $\mathfrak{D}_\mathfrak{I}=\R^{D\times D}\cap\mathfrak{D}\left({ \mathbb{H}}\right)$ contains pure states (all pure redit states, as stated). 
This is what makes it an appropriate invariant subspace in which to search for a PRE.  A simple way to engineer such an invariant subspace is to choose a Lindbladian  which has a real-valued matrix representation in a basis where it is acting on column-stacked density matrix elements.  Interestingly, this will lead to the condition 
${\bf L}_{{\mathfrak{I}_{0},\mathfrak{R}_{0}}}=0$, which means that 
$\mathfrak{R}_{0}$ will also be invariant. 
However, for the case of $D>2$, the latter space does not contain any pure states and is, therefore, not capable of supporting a PRE. This way to engineer
$\mathfrak{D}_\mathfrak{I}=\R^{D\times D}\cap\mathfrak{D}\left({ \mathbb{H}}\right)$ is not, however, the only way --- we will provide an example in the following subsection for which $\mathfrak{R}_{0}$ is not invariant.

\begin{table}

\begin{indented}
\lineup
\item[]\begin{tabular}{lllllll}
\br       
&\centre{5}{Dimension, $D$}\\     
\ns   \ns
&\crule{5} \\            
&$2$&$3$&$4$&$5$&$6$\cr 
\mr
{\rm redit}&2  & 4&7&11&16\cr
{\rm qudit}&2 & 5&10&17&26  \cr 
\br
\end{tabular}
\end{indented}
\caption{\label{table1}The heuristically argued minimum number of PRE members, $K$, required for the number of parameters to equal or exceed the number of constraints, is provided for small dimension, $D$. The comparison being made is between MEs that have real-valued density matrices as an invariant subspace (redits) and those that do not (qudits).  The parameter counting heuristic suggests that a solution of \erf{jumpCond} may be possible for $K$ equalling or exceeding these values.  We are considering MEs that are generic, apart from the presence of the invariant subspace symmetry. 
}
\end{table}

\subsection{Qubit examples}
\subsubsection{Resonance fluorescence}
\label{RF}
We explore further the nature of PREs possessing an invariant subspace symmetry --- specifically,
that of a real subspace $\mathfrak{D}_\mathfrak{I}=\R^{D\times D}\cap\mathfrak{D}\left({ \mathbb{H}}\right)$, discussed in \srf{realRho} --- by re-examining the $D=2$ (qubit) resonance fluorescence system that was introduced in \srf{RFa}, following \cite{karasik2011many,karasik2011tracking}.
Given the Hamiltonian, $\hat{H}=\frac{\Omega}{2}\hat{\sigma}_{x}=i\Omega\left(\ketbra{0}{1}-\ketbra{1}{0}\right)/2$, and the single Lindblad jump operator, $\hat{c}=\sqrt{\gamma}\ketbra{0}{1}$, \blk it is clear that a density matrix that is initially real-valued will stay real-valued.  (We remind the reader that the rebit plane is the $y$-$z$ plane of the Bloch ball.)  The steady state is given by $\pmb{x}_{{\rm ss}}=(0,2\gamma\Omega,-\gamma^2)^{{\rm T}}/(\gamma^2+2\Omega^2)$.

As described in \srf{defs}, the right-eigenvectors of ${\bf L}_{0}$ are used to identify invariant subspaces of interest. \blk The three right-eigenvectors of ${\bf L}_{0}$ (see \erf{LBlockResFlu}) are: $ \pmb{e}_1=(1,0,0)^{{\rm T}}$ and $ \pmb{e}_{\pm}=(0,\gamma\pm\sqrt{\gamma^2-16\Omega^2},4\Omega)^{{\rm T}}$ (the latter two being unnormalized).
When all eigenvectors are real ($\gamma^2-16\Omega^2\geq 0$), each eigenvector gives a one-dimensional invariant subspace.  When $\gamma^2-16\Omega^2<0$, the real and imaginary components of $\pmb{e}_{+}$ (equivalently $ \pmb{e}_{-}$) are used to form a two-dimensional subspace,  and the one-dimensional subspace corresponding to $ \pmb{e}_1$ remains also.  
The structure of the eigenspaces is such that the space formed by $ \pmb{e}_{1}$ is always orthogonal to the other spaces (be they one- or two-dimensional). However, the other one-dimensional spaces (when they exist) are not respectively orthogonal, but together span 
the space orthogonal to $\pmb{e}_{1}$.

The connection to the space $\mathfrak{U}_{0}$ (being a displaced $3$-ball 
parameterized by $u,v,w$\blk) and $\mathfrak{I}_{0}$ is as follows. 
When $\gamma^2-16\Omega^2\geq 0$, there are three one-dimensional $\mathfrak{I}_{0}$ spaces that are defined by the following directions: the $u$-axis (due to  $\pmb{e}_{1}$) and two other rays lying in the $u=0$ great disc (due to $\pmb{e}_{\pm}$).  When $\gamma^2-16\Omega^2< 0$, $\mathfrak{I}_{0}$ is either (along) the $u$-axis or the full $u=0$ great disc.  If $\mathfrak{I}_{0}$ is taken to be the $u$-axis, there is a duality, in that $\mathfrak{R}_{0}$ itself is invariant (being the $u=0$ great disc).  This could be inferred from \erf{LBlockResFlu} as it has the form of \erf{LBlock}, but with $ {\bf L}_{\mathfrak{I}_{0},\mathfrak{R}_{0}}=0$. 
This duality, $ {\bf L}_{\mathfrak{I}_{0},\mathfrak{R}_{0}}= 0$, is not present in the case 
when both $\gamma^2-16\Omega^2\geq 0$ and the invariant subspace $\mathfrak{I}_{0}$ is chosen as one of the two rays lying in the $u=0$ great disc.  The non-orthogonality of $\pmb{e}_{\pm}$ ensures that a state initialized in $\mathfrak{R}_{0}$ will not, in general, remain in $\mathfrak{R}_{0}$ 
The reader should remember that $\mathfrak{U}_{0}$ (in which the vectors $\pmb{u}$ live) is not the Bloch ball (centered at the origin), but rather a unit ball with origin $\pmb{x}_{{\rm ss}}$.

Recalling our requirement that $\mathfrak{I}_{0}$ be of dimension at least $D-1=1$ and that it contain pure states (when mapped to density matrices via \erf{rhoReduced}), we see that all of the eigenspaces discussed above are interesting, and we can search for PREs contained fully in each of them, respectively.  Note that {\it all} the extreme states of $\mathfrak{U}_{0}$ correspond to pure states, a feature unique to $D=2$. 

First, we consider the one-dimensional $\mathfrak{I}_{0}$ subspaces in the context of PREs.  
When $\gamma^2-16\Omega^2\geq 0$ each of the three $\mathfrak{I}_{0}$ meets the surface of $\mathfrak{U}_0$ in two locations, 
corresponding to two pure states on the Bloch sphere. Remembering that the PRE must reside in $\mathfrak{I}_{0}$ (as per \erf{invSubspaceFrak}), we see that it is only feasible for PREs with $K=2$, and each extremal point in $\mathfrak{I}_{0}$ must be an ensemble member.  

The case of $D = K=2$ was treated analytically in Ref.~\cite{karasik2011many}, with expressions describing the PRE resulting; these have been provided in \erfs{analyticPRE1}{analyticPRE2} \blk .  Translating these results to the space $\mathfrak{U}_{0}$ --- which we have defined in our work --- the PRE states are
\beq
 \pmb{u}_{k}=-\eta_{j}\blk (-1)^{k} \pmb{e}_{j},
\label{preUSpace2D}
\eeq  
where $k=\{1,2\}$ labels the ensemble member, $j$ labels the real eigenvectors, $\pmb{e}$, of ${\bf L}_{0}$ and $\eta_{j}$ is a scalar that depends on $ \pmb{x}_{{\rm ss}}$, $\pmb{e}_{j}$, and $k$.  The important point for our purposes is that the PRE states in $\mathfrak{U}_{0}$ are in the direction of $\pmb{e}_{j}$ --- in other words, {\it all} $K=D=2$ PREs lie in $\mathfrak{I}_{0}$ and each one-dimensional $\mathfrak{I}_{0}$ supports a PRE (via its corresponding real eigenvector).
\blk
Each of the $K=D=2$ PREs have been displayed in
\frf{2KPREs}, with subplots (a),(b) associated with $ \pmb{e}_{\pm}$ and subplot (c) with $ \pmb{e}_{1}$.
In addition to plotting the PREs on the Bloch ball, subplot (d) shows the PRE of subplot (a) in the $u=0$ subspace of $\mathfrak{U}_0$. 
\blk
When $\gamma^2-16\Omega^2<0$, the one-dimensional spaces corresponding to eigenvectors $ \pmb{e}_{\pm}$ are no longer available and there exists only a single $K=2$ PRE which is attributable to the $ \pmb{e}_{1}$ eigenspace.

%

For $K=3$, the $\mathfrak{D}_\mathfrak{I}$ arising from the three one-dimensional $\mathfrak{I}_{0}$ (that pierce the $\mathfrak{U}_0$ \blk ball at only two points) cannot hold PREs, so it is natural to look at the two-dimensional $\mathfrak{I}_{0}$ subspace (being the $u=0$ great disc).  A complete discussion of $K=3$ is given in Ref.~\cite{karasik2011tracking} for the interested reader --- in this paper we just wish to point out the symmetry that exists in some but not all of the $K=3$ PREs.  For example, 
\frf{3KPREs}(a),(c),(d),(f)~\footnote{As per \ref{footRef2}, but \frf{3KPREs}(a)-(f) \blk originally appeared in Ref.~\cite{karasik2011tracking} (with the same parameters).} 
all contain PREs with $u=0$ (corresponding to $x=0$ in the Bloch ball figures). \blk   That is, they reside in the $\mathfrak{D}_\mathfrak{I}$ pertaining to the two-dimensional $\mathfrak{I}_{0}$ discussed above.  
Importantly, there are PREs that do {\it not} possess the invariant subspace symmetry (see \frf{3KPREs} subplots (b) and (e)).  This illustrates the fact that by searching for PREs in a reduced space [$\mathfrak{D}_\mathfrak{I}$ compared with the total space $\mathfrak{D}\left({ \mathbb{H}}\right)$], one may not find all PREs 
for a given $K$.  
This will also be a feature when we consider a different type of symmetry (Wigner symmetries) in the next section.  It is important to note that, following Ref.~\cite{karasik2011many}, we have limited to {\it cyclic} 
PREs when illustrating our logical points.  If non-cyclic PREs exist (necessarily having $K>2$), some of them may well possess the invariant subspace symmetry; a search for them would be significantly simplified utilizing the methods of this section. 


\begin{figure}[t]
\centering
 \includegraphics[width=0.9\textwidth]{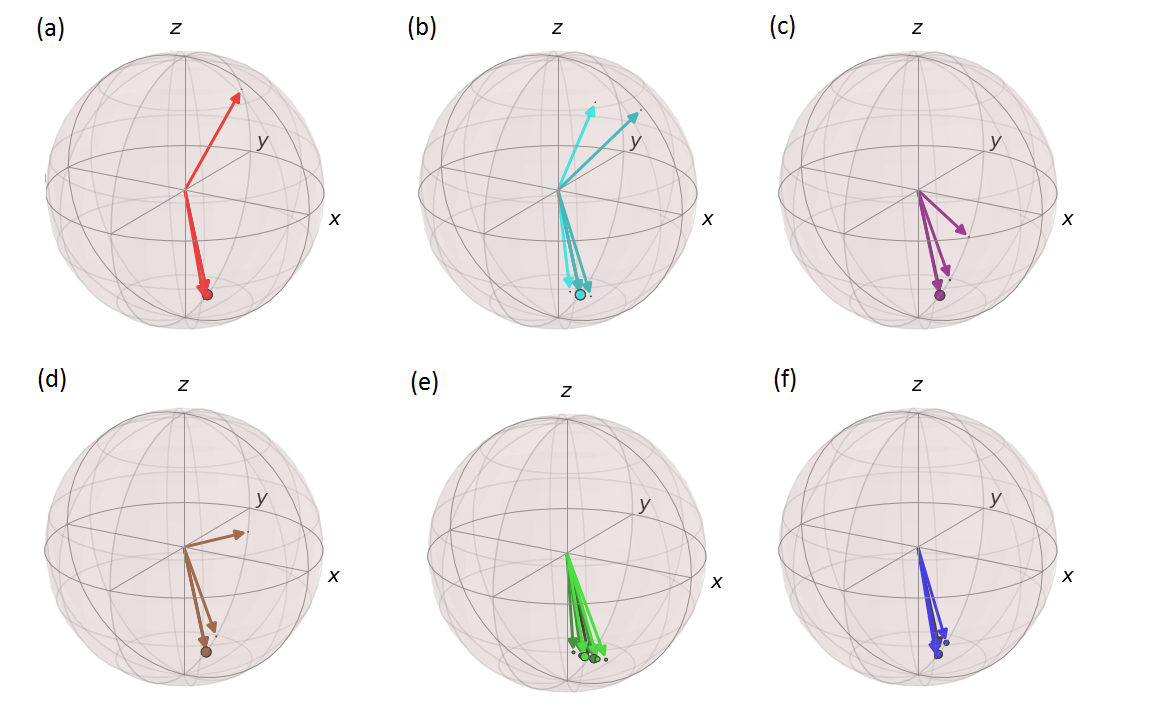}
\caption{Coloured arrows show Bloch vectors for all eight $K=3$ PREs of the ME of \erf{LBlockResFlu}. Two PREs are shown in each of subplot (b) and (e) using different shades of colour.  Other details are as per \frf{2KPREs}. 
}
\label{3KPREs}
\vspace{-5mm}

\end{figure}

\subsubsection{Absorption and emission ME}
\label{invSubAbDis}
Another qubit ME that possesses much symmetry is that which models incoherent emission and absorption.  That is, we set $\hat{H}=0$, but add a second decoherence channel, as compared with \srf{RF}, so that $\hat{c}_{1}=\sqrt{\gamma_{-}}\ketbra{0}{1}$ and $\hat{c}_{2}=\sqrt{\gamma_{+}}\ketbra{1}{0}$.  This ME was briefly considered, in the context of PREs, in Ref.~\cite{1367-2630-16-6-063028}; the reader is also referred to the application discussed in Ref.~\cite{PhysRevLett.108.170501} as evidence of its relevance to quantum information.  In the Pauli basis we obtain
\beq
{\bf L}_{0}=
-\gamma_{\Sigma}\left[\begin{array}{ccc}
   1/2       &0&0   \\
    0     			 &1/2&0\\
    0				&0	     &1
\end{array}\right],
\quad  \pmb{b}=\left[
\begin{array}{c}
   0         \\
    0        \\
    \gamma_{\Delta}				
\end{array}\right],
\label{absDis}
\eeq
where $\gamma_{\Sigma}=\gamma_{-}+\gamma_{+}$ and $\gamma_{\Delta}=\gamma_{+}-\gamma_{-}$.  Without loss of generality we assume $\gamma_\Delta \leq 0$~\footnote{Because the ME is invariant under 
$\ket{0}\leftrightarrow \ket{1}$ and $\gamma_{+}\leftrightarrow\gamma_{-}$, the results for $\gamma_\Delta \geq 0$ can be simply obtained from those for $\gamma_\Delta \leq 0$. 
If a ME, parameterized by $\{\gamma_{+},\gamma_{-}\}$, possesses a PRE with $K$ members $\{x_{k},y_{k},z_{k}\}_{k=1}^{K}$, then the ME parameterized by $\{\gamma_{-},\gamma_{+}\}$ will possess a PRE of the form $\{x_{k},-y_{k},-z_{k}\}_{k=1}^{K}$.}.
The steady state is $\pmb{x}_{{\rm ss}}=(0,0,\gamma_{\Delta}/\gamma_{\Sigma})^{{\rm T}}$, which lies on the $z$-axis of the Bloch ball.  The three right-eigenvectors of ${\bf L}_{0}$ (all of which are real) are $\pmb{e}_{1,2,3}=\{  \ro{1,0,0}^{{\rm T }},\ro{0,1,0}^{{\rm T }},\ro{0,0,1}^{{\rm T }}\}$ having eigenvalues $-\gamma_{\Sigma}\{1/2,1/2,1\}$ respectively.  An important difference from the resonance fluorescence case is that there is a degeneracy of eigenvalues, the consequence of which is that {\it every} diameter of the $w=0$ disc of the $3$-ball $\mathfrak{U}_{0}$ is an invariant one-dimensional subspace, $\mathfrak{I}_{0}$.  Of course, the portion of the $w$-axis within $\mathfrak{U}_{0}$ is also an invariant one-dimensional subspace.  The case of $K=2$ is discussed in Ref.~\cite{1367-2630-16-6-063028}, and it is indeed the case that each of the invariant subspaces contains a $K=2$ PRE.  An infinite number of PREs are parametrized by the azimuthal angle in the $w=0$ disc, and there is one further PRE consisting of the two points where the $w$-axis and $\mathfrak{U}_{0}$ intersect.  In summary, each $\mathfrak{I}_{0}$ space contains a $K=2$ PRE and there are no $K=2$ PREs that do not possess the invariant subspace symmetry --- this is expected as per the discussion below \erf{preUSpace2D}.\blk  

$K=3$ PREs were not investigated in Ref.~\cite{1367-2630-16-6-063028}, but it is of interest for us to search for them, making use of the discussed invariant subspace PRE symmetry technique.  For simplicity, we once again limit the investigation to cyclic PREs.  Of course the one-dimensional $\mathfrak{I}_{0}$ are not sufficient for $K=3$ PREs as they only correspond to two pure states.  Instead, we form two different two-dimensional $\mathfrak{I}_{0}$ within the $3$-ball $\mathfrak{U}_{0}$: firstly, the $w=0$ disc and, secondly, the $v=0$ great disc.  Within the first subspace there are no cyclic $K=3$ PREs. This is an analytic result, as the constraints imposed by \erf{jumpCond} can be shown to be algebraically inconsistent for arbitrary, but non-zero, $\gamma_{\pm}$~\footnote{\label{refFoot}A straightforward way to verify this is to formulate the constraint equations in terms of the  Bloch vectors (rather than the state vectors of \erf{jumpCond}) then, due to the symmetry of the system, choose say $v_{1}=0$.  By inspection, this will lead to $v_{2}=v_{3}=0$ also.  This is a contradiction as it excludes the possibility of three distinct states in the ensemble.  Resultantly, we conclude that there are no such PREs contained in the $w=0$ disc.}. Within the 
$v=0$ great disc, our numeric \blk results depend on the parameter regime.  For 
$\gamma_{+}>\varepsilon\gamma_{-}$, with $\varepsilon \approx 1/18$,  
no PREs are found (recall that we restrict to $\gamma_{-} \geq \gamma_+$). 
But for $\gamma_+ < \varepsilon\gamma_{-}$, 
two cyclic $K=3$ PREs exist. Examples of these \blk are shown in \frf{daryPRE}.

The results pertaining to the $v=0$ great disc are numeric in that we specify the ratio $\gamma_{+}/\gamma_{-}$ before applying Gr{\"o}bner basis techniques~\cite{cox2006using}
either of a algebraic (using the computational algebra software MAGMA~\cite{bosma1997magma}) or numeric nature (MATHEMATICA) to solve the polynomial system.  
That is, we can be sure of the existence of PREs only at discrete $\gamma_{+}/\gamma_{-}$ ratios.  However, by conducting a fine-grained search (we stepped in $0.001$ increments) we can develop a clear picture~\footnote{\,\,Fortunately, the reduction of the size of the polynomial system, due to the invariant subspace symmetry being utilized, made the analysis of the system simple enough for MATHEMATICA to carry out very quickly.  This facilitated the process of testing at thousands of different $\gamma_{+}/\gamma_{-}$ ratios.  Results were selectively compared with MAGMA for consistency.}.

In fact, because of the ME having no preference in the $u$-$v$ plane ($x$-$y$ plane in the Bloch ball), {\it any} plane containing all of the $w$-axis is actually a two-dimensional $\mathfrak{I}_{0}$ and possesses PREs corresponding to azimuthal rotations of the $v=0$ disc PREs~\footnote{\,\,Technically there are 4 PREs in the $v=0$ great disc, but two of them are obtained from the other two by azimuthal rotation by $\pi$.}.
That is, we can obtain an infinite number of  PREs (forming a `family') by rotation of each member of the PRE about the $w$-axis --- this is in complete analogy with the family of $K=2$ PREs that are related by rotation.  In this section, these families appear somewhat {\em ad hoc}; in the next section, the theory of such PRE families will be explicated, with their origin attributed to a symmetry different from the invariant subspace symmetry.  Following a general theoretical development, we will return to our qubit examples.

\begin{figure} [t]
\centering
\includegraphics[width=0.35\textwidth]{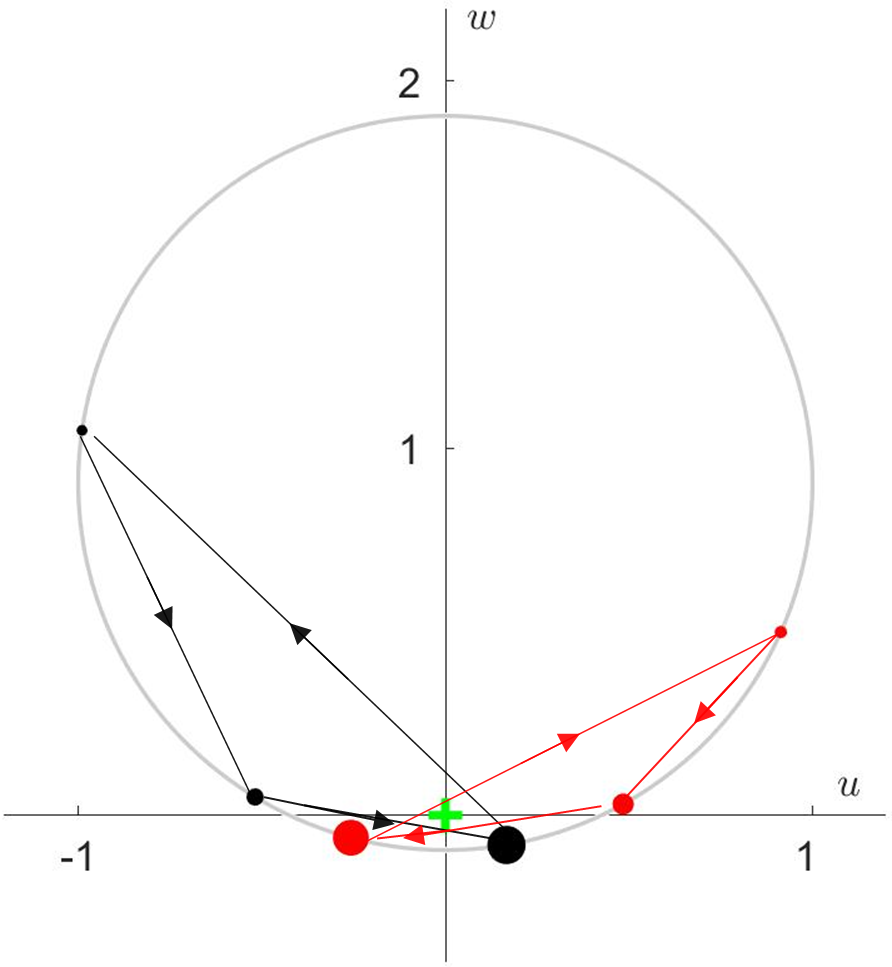} 
\vspace{-2mm}
\caption{Two cyclic $K=3$ PREs are shown for the ME of \erf{absDis} --- each is comprised of dots of a single colour, either red or black.  As they lie on the $v=0$ great circle, only that great disc of $\mathfrak{U}_{0}$ is displayed. 
The relative area of the dots indicates the relative occupation probability amongst the ensemble members of each ensemble respectively.  The direction of cycling within each PRE is shown by the arrows.  The steady state location is shown by the green cross, which, by definition, lies at the origin in the $u,v,w$ coordinate system. \blk The parameter choice is $\gamma_{+}=0.05\gamma_{-}$. 
}
\label{daryPRE}
\end{figure}

For completeness, a search for $K=3$ cyclic PREs for the ME of \erf{absDis} can be considered.  However, by inspection of the constraint equations, it quickly becomes analytically apparent that no further PREs beyond those contained within $\mathfrak{I}_{0}$ are possible.
The demonstration of this is as per \ref{refFoot}, where, once again, the symmetry of the system allows a simplification that quickly leads to the conclusion that a PRE must lie on a great disc containing the $w$-axis.

\section{Wigner invariance of ${\cal L}$}
\label{uniSymm}

Wigner transformations act  on Hilbert space rays in a way that preserves the Hilbert space inner product, $|\braket{\psi_{1}}{\psi_{2}}|$. Their action is consequently well defined upon pure state projectors, and we denote this as ${\cal T}\ketbra{\psi}{\psi}$. \blk  Wigner showed that such transformations are either unitary (and so linear in their action on Hilbert space) or antiunitary (and so antilinear)~\cite{wigner1,wigner1959group}.  In this section, we consider those Wigner transformations which leave the Lindbladian, ${\cal L}$, invariant:
\beq
{\cal T}^{-1}{\cal L}\ {\cal T}={\cal L},
\label{Uinv}
\eeq
and term these `Wigner symmetries'.

Using the unitary/antiunitary properties of ${\cal T}$ and the block structure of ${\bf L}$ implied by \erf{lindVecReduced} we can gain further insight into \erf{Uinv}, which reduces to 
\bqa
 {\bf T}_{0}^{-1}{\bf L}_{0} {\bf T}_{0}={\bf L}_{0}\quad\quad{\rm and}
\label{wigSymm1}\\
 {\bf T}_{0} \pmb{b}=\pmb{b},
\label{wigSymm2}
\eqa
where ${\bf T}_{0}$, analogously to ${\bf L}_{0}$,  is the restriction of the matrix representation, ${\bf T}$, of the superoperator ${\cal T}$ to traceless Hermitian matrices.  Given that the ME has, by assumption, a unique steady-state, we can use \erfs{wigSymm1}{wigSymm2} and \erf{steadyState} 
to infer that the steady is invariant under the action of ${\cal T}$,
\beq
{\bf T}_{0}\pmb{x}_{{\rm ss}}=\pmb{x}_{{\rm ss}}.
\label{ssWig}
\eeq
Together, \erf{wigSymm1} and \erf{ssWig} provide a useful formulation of \erf{Uinv}.
\blk

\subsection{Wigner-symmetric families of PREs} \label{WIFs}
By using the Wigner symmetry of \erf{Uinv} in \erf{jumpCond}
we can investigate its significance for PREs.  After utilizing the symmetry and then premultiplying by ${\cal T}$ we obtain
\beq
  \forall k, \ {\cal L} {\cal T}\ket{\phi_k}\bra{\phi_k}= \sum_{j=1}^K \kappa_{jk} 
 \left(
{\cal T}\ket{\phi_j} \bra{\phi_j}-{\cal T}\ket{\phi_k} \bra{\phi_k} 
 \right).
\label{jumpCondSymm} 
\eeq
Note that the $\kappa_{jk}$, because they are real valued, commute with all ${\cal T}$, including antiunitary transformations.
It is clear that a solution to \erf{jumpCond} can be used to generate new PREs under the action of ${\cal T}$.  Given how computationally difficult it is to find a PRE, this is very valuable.  We also note that the set of all ${\cal T}$ satisfying \erf{Uinv} forms a group, $\mathbb{G}$. This is apparent since: the identity 
${\cal I}$ satisfies \erf{Uinv}, every Wigner transformation 
${\cal T}$ has an inverse ${\cal T}^{-1}$ that is also a Wigner transformation, and if 
$ {\cal T}_{1}$  and ${\cal T}_{2}$ satisfy \erf{Uinv} then so does 
 $ {\cal T}= {\cal T}_{2}\ {\cal T}_{1}$. 
If $\mathbb{G}$ (or a subgroup of $\mathbb{G}$) is a Lie group then there is an interesting consequence, 
namely that an infinite number of PREs can be found irrespective of whether or not the constraining polynomial system (defined by \erf{jumpCond}) is square. 
(Note that a Lie group could not include antiunitary symmetries, which are necessarily discrete.) 

\subsection{Wigner-symmetric PREs}

In addition to being used to generate families of PREs, as discussed above, the Wigner symmetry described by a group $\mathbb{G}$ can also be applied to make it easier to find individual PREs.  To do so we first introduce the following notation.  Let $P$ be a permutation of $\{1...K\}$ with $k^{\prime}=P(k)$.  Then we define a PRE as having the Wigner invariance for some ${\cal T}\in\mathbb{G}$ iff $\exists$ permutation $P$ such that
\bqa
\forall k, \ {\cal T}\ket{\phi_k}\bra{\phi_k} = \ket{\phi_{k^{\prime}}}\bra{\phi_{k^{\prime}}}\quad\quad{\rm and}
\label{induced}\\
 \kappa_{j^{\prime}k^{\prime}} =\kappa _{jk}.
\label{inducedMap}
\eqa
If this holds, then the action of ${\cal T}$  (or ${\cal T}^{-1}$) on any ensemble member generates an existing member, which could be itself ($k^{\prime}=k$) or a different member ($k^{\prime}\neq k$).  It is possible that \erfs{induced}{inducedMap} are satisfied simultaneously, for more than one non-identity Wigner symmetry in $\mathbb{G}$. Such elements together would form a subgroup $\mathbb{K}$ of the group $\mathbb{G}$. 
The different elements in $\mathbb{K}$ 
may require different permutations $P$ in \erfs{induced}{inducedMap}. 
However, the number of distinct permutations is, of course, finite for a PRE of finite size $K$.
As \erf{induced} places constraints on the ensemble members, the cardinality of the group  $\mathbb{K}$ will, in general, be smaller than that of $\mathbb{G}$ (the application of symmetries in addition to those of $\mathbb{K}$ would lead to inconsistencies).

The consequence of both ${\cal L}$ and the PRE possessing the Wigner symmetry [\erf{Uinv} and \erfs{induced}{inducedMap}, respectively] is that the $k$th constraint of \erf{jumpCondSymm},
\beq
 {\cal L}{\cal T}\ket{\phi_k}\bra{\phi_k} = 
 \sum_{j=1}^K \kappa_{jk} 
 \left({\cal T}\ket{\phi_j} \bra{\phi_j}-{\cal T}\ket{\phi_k} \bra{\phi_k} \right),\label{PREsymmA}
 \eeq
 implies that the $k^{\prime}$th is satisfied also:
\beq 
{\cal L}\ket{\phi_{k^{\prime}}}\bra{\phi_{k^{\prime}}} = 
 \sum_{j^{\prime}=1}^K \kappa_{j^{\prime}k^{\prime}} 
 \left(\ket{\phi_{j^{\prime}}} \bra{\phi_{j^{\prime}}}-\ket{\phi_{k^{\prime}}} \bra{\phi_{k^{\prime}}} \right).
\label{PREsymmB}
\eeq
If we define an equivalence relation, $\sim$, amongst ensemble members in the presence of the symmetry ${\cal T}$ as 
\beq
\ket{\phi_k}\sim\ket{\phi_{k^{\prime}}} \quad {\rm iff} 
\quad 
\exists \ \  {\cal T}\in\mathbb{K}: \ 
\ket{\phi_{k^{\prime}}}\bra{\phi_{k^{\prime}}}=
{\cal T}\ket{\phi_{k}}\bra{\phi_{k}},
\label{equivClass}
\eeq
then the constraints on only one element of each equivalence class, $[\ket{\phi_k}]_{\sim}$, 
need to be tested as the remainder are implied.  A PRE being Wigner-symmetric is consistent with the invariance with which we began in \erf{Uinv}~\footnote{\,\,We show this as follows: \blk starting from \erf{PREsymmB}, the PRE symmetry implies \erf{PREsymmA} which can be rearranged to give ${\cal T}^{-1}{\cal L}\ {\cal T}\ket{\phi_{k}}\bra{\phi_{k}}= \sum_{j=1}^K \kappa_{jk} 
 \left(\ket{\phi_j} \bra{\phi_j}-\ket{\phi_k} \bra{\phi_k}\right)$.  For this to agree with \erf{jumpCond} we require that  $ \forall k,\ {\cal T}^{-1}{\cal L}\ {\cal T}\ket{\phi_{k}}\bra{\phi_{k}}= { \cal L}\ket{\phi_{k}}\bra{\phi_{k}}$. 
This is clearly compatible with ${\cal T}^{-1}{\cal L}\ {\cal T}={\cal L}$ and the PRE symmetry is thus consistent with the Lindbladian invariance.}. 
The equivalence class $[\ket{\phi_k}]_{\sim}$ will have more than two members if the different  elements in $\mathbb{K}$ require different permutations $P$ in order to satisfy \erfs{induced}{inducedMap}.  As an example, this situation will arise if the period of the elements of $\mathbb{K}$ is greater than two (that is, ${\cal T}^{2}\neq \mathbb{1}$).

\subsection{Qubit examples}
\subsubsection{Resonance fluorescence}
We can, once again, use the resonance fluorescence qubit ME described by \erf{LBlockResFlu}, this time to illustrate the Wigner symmetry.   It is invariant (in the sense of \erfs{wigSymm1}{wigSymm2}) under the antiunitary transformation ${\bf T}_{0} = {\rm diag}(-1,1,1)$, defined here using the Pauli operator basis $\{\sigma_{x},\sigma_{y},\sigma_{z}\}$, where we have restricted to the traceless Hermitian operators.  The steady-state is also invariant under the symmetry's action, as required by \erf{ssWig}.  Clearly this Wigner symmetry, which takes $\sigma_{x}\rightarrow -\sigma_{x}$, is a $\mathbb{Z}_{2}$ symmetry, meaning ${\cal T}^2=\mathbb{1}$.

All three of the $K=2$ PREs obey the Wigner symmetry, ${\cal T}\ketbra{\phi_{k}}{\phi_{k}}=\ketbra{\phi_{k^{\prime}}}{\phi_{k^{\prime}}}$. For two of them --- see \frf{2KPREs}(a),(b) --- the permutation is trivial and each ensemble member is mapped to itself ($k=k^{\prime}$).  These PREs necessarily lie wholly on the $x=0$ great disc which is the ${\cal L}$-invariant subspace attributable to $\pmb{e}_{\pm}$.  The remaining $K=2$ PRE, which lies in the ${\cal L}$-invariant subspace attributable to $\pmb{e}_{1}$ --- see  \frf{2KPREs}(c) --- has a non-trivial permutation associated with ${\cal T}$; the ensemble members are mapped to each other.

Regarding the $K=3$ cyclic PREs, four of the eight possess the Wigner symmetry (see \frf{3KPREs}(a),(c)(d),(f)), but only in the trivial manner just described, lying on the $x=0$ great disc.    
Those PREs that are not Wigner symmetric come in pairs, such that one PRE in the pair is obtained from the other under the action of ${\cal T}$, in the manner 
discussed in \srf{WIFs}, as seen in \frf{3KPREs}(b),(e).  
The fact that some of the PREs in this example do not possess the Wigner symmetry highlights that, similarly to the case of invariant subspaces of ${\cal L}$, discussed in the previous subsection, by imposing PRE Wigner symmetry we may not find the entire solution set. 

It is not a coincidence that the Wigner-symmetric $K=3$ cyclic PREs of the resonance fluorescence ME, that are found in Ref.~\cite{karasik2011tracking}, have the symmetry present in only a trivial form, with each ensemble member being mapped to itself.  The specific Wigner-symmetry in question is ${\bf T} = {\rm diag}(-1,1,1,1)$, so that a $K=3$ PRE possessing the symmetry, in a non-trivial manner, would have to have two ensemble members of the form $\ket{\phi_1}\bra{\phi_1}\equiv (x,y,z)$ and $\ket{\phi_2}\bra{\phi_2}\equiv (-x,y,z)$ for $x\neq 0$, and a third ensemble member as $\ket{\phi_3}\bra{\phi_3}\equiv (0,y_{3},z_{3})$.  Additionally, according to \erf{inducedMap}, we must have $\kappa_{1,3}=\kappa_{2,3}$ and $\kappa_{3,1}=\kappa_{3,2}$.  This is inconsistent with the assumed cyclic nature of the PRE and, as a consequence, rules out their existence. 

It is feasible to look for non-cyclic $K=3$ PREs possessing the Wigner symmetry --- the PRE structure just described collapses the polynomial system greatly. However, a difference to the previously discussed numerical searches is that the polynomial system is no longer square.  The additional potential transitions beyond cyclicity provide extra variables, which then exceed, in number, the constraint equations (in this case, by one).  Generically, one can then expect a positive dimension solution set (here of dimension one), before imposing that we require real-valued solutions and positive transition rates.  To avoid this difficulty, we run repeated numerical tests at discrete values of a chosen variable ($100$ evenly spaced values of $x_{1}$), 
thus reducing the number of parameters to be equal to the number of constraints.  This collapses the problem to a set of square polynomial systems.  Furthermore, we also carry out this procedure for a series of $200$ different MEs, parameterized by the value of $\Omega/\gamma$ (varied between $0$ and $10$).
\blk
Despite the fact that the obstacle of the previous paragraph is avoided (by allowing non-cyclicity), no {\it non-trivial} Wigner symmetric PREs were found. We can be reasonably confident that none exist, but not certain, due to the numerical nature of the search. \blk The reader should not think that the apparent non-existence of non-trivial PREs with $K>2$ in this example indicates that none are possible for a qubit. \blk In the ME example of \erf{absDis}, Wigner symmetric PREs for arbitrarily large $K$ exist, and will be presented in \srf{absDisCombo}.

\subsubsection{Absorption and emission ME}
We now return to the ME of \erf{absDis} that describes incoherent absorption and emission, and consider Wigner symmetries.  It is not hard to identify that it has the orthogonal Lie group, $\mathbb{O}(2)$ (consisting of rotations and reflections), acting on the first two coordinates, as a Wigner symmetry.  That is, 
\beq
{\bf T}_{0} =\left[
\begin{array}{c|c}
R(2)      &0   \\
\hline
    0     			 &1\\
\end{array}\right],\quad \forall \ R(2)\in\mathbb{O}(2).
\label{genT}
\eeq
For clarity, $R(2)$ is a $2\times 2$ matrix representation of the group $\mathbb{O}(2)$.  Note that although an inversion of the $z$-coordinate would also satisfy \erf{wigSymm1}, it does not satisfy \erf{wigSymm2} (unless $\gamma_{+}=\gamma_{-}$); thus, the lower right element of ${\bf T}_{0}$ must be unity.   
\blk

It is therefore expected that families of PREs related by the action of ${\bf T}_{0}$ will be obtained.  On the Bloch sphere, this means that for a given PRE, we expect a second PRE to exist that is related to the first by rotation about the $z$-axis and/or the reflection about a plane containing the entire \blk $z$-axis.  As $\mathbb{O}(2)$ is a Lie group, an infinite number of PREs (some of which may be indistinct) will belong to each family.  For $K=2$, the PREs belong to two families: those contained in the $z=z_{{\rm ss}}$ plane of the Bloch ball and the PRE comprised of the poles of the Bloch ball.  The latter PRE has only one PRE in its family as it is mapped to itself.  As expected, the actual $K=2$ PREs conform to these predictions.
A description of cyclic $K=3$ PREs was given in \srf{invSubAbDis}, and we can also now understand the origin of those families of PREs (that exist for 
$\gamma_{+}\lesssim\gamma_{-}/18$): \blk
they are being generated under the action of ${\bf T}_{0}$, just as for the $K=2$ PREs.

We now consider Wigner-symmetric PREs; that is, each of the members of a given ensemble must be mapped to one another under the action of ${\cal T}$, with the transition rates also related by \erf{inducedMap}.  The PRE contained on the $z$-axis is a Wigner symmetric PRE, but only under the trivial permutation in which each member is mapped to itself.  As the transition rates between ensemble members are different when $\gamma_{+}\neq\gamma_{-}$  (when $\gamma_{+}=\gamma_{-}$ the ME has the additional Wigner symmetry under reflection in the $z=0$ plane) 
it is not possible for this PRE to have a $\mathbb{Z}_{2}$ symmetry.  For $K=2$, we require that ${\bf T}_{0}^{2}=\mathbb{1}$, which implies that $R(2)$ is either a rotation by $\pi$ or a reflection. Note that $K=2$ ensembles whose states are mapped to each other under reflection, but are not antipodal, cannot form PREs as their ensemble average will not lie on the $z$-axis. \blk  Wigner symmetric $K=2$ PREs are, therefore, those that consist of the antipodal points of the intersection of the $z=z_{{\rm ss}}$ plane and the Bloch ball and, additionally, have each member occupied with equal probability.  Thus, the family of $K=2$ PREs lying in the $z=z_{{\rm ss}}$ Bloch plane are Wigner symmetric.     

As for the cyclic $K=3$ PREs, these are not Wigner symmetric --- they possess neither a ${\bf T}_{0}^{2}=\mathbb{1}$ (with the third ensemble member mapped to itself) nor a ${\bf T}_{0}^{3}=\mathbb{1}$ symmetry (which can be seen as this would require them to be mapped outside of the plane containing the $z$-axis under the action of ${\bf T}_{0}$).  The theory to investigate some larger ($K> 3$) and more complicated (non-cyclic) PREs, for this ME, is now in place, but we postpone this to the following section, as the combination of symmetries makes finding some of them particularly easy.

\blk
\section{Combining symmetries}
\label{combo}

The reader will have noticed that many of the example PREs thus far presented have possessed both the Wigner symmetry and the invariant subspace symmetry.  
In this section we further explore the simultaneous existence of these symmetries at a ME and PRE level.

The requirement that a ME possess the invariant subspace symmetry of \srf{invSubspaceSec} in addition to the Wigner symmetry of \erf{Uinv}
can be formulated as requiring \erf{wigSymm1} to be satisfied when ${\bf L}_{0}$ has the block form given in \erf{LBlock}.  This needs to be supplemented by \erf{ssWig} as it is non-trivial that the steady-state will possess the Wigner symmetry.
Given the satisfaction of these constraints, 
then the results of \srf{invSubspaceSec} and \srf{uniSymm} can clearly be applied to search for PREs, for the ME in question, that possess either one of the two symmetries.  

The obvious next question is, under what conditions can we look for PREs that possess both symmetries {\it jointly}?  This would allow us to leverage both symmetries and greatly simplify the finding of (jointly symmetric) PREs.  
To investigate, we give ${\bf T}_0$ the block structure induced by that of ${\bf L}_0$ (see below \erf{LBlock} for further details)
\beq
{\bf T}_0=
\left[
    \begin{array}{c|c}
      {\bf T}_{\mathfrak{I}_{0} } & {\bf T}_{\mathfrak{I}_{0},\mathfrak{R}_{0} }\\
      \hline
     {\bf T}_{\mathfrak{R}_{0} ,\mathfrak{I}_{0}} & {\bf T}_{\mathfrak{R}_{0}}
    \end{array}
    \right].
\eeq

Firstly, from \erf{induced}, in the basis for which ${\bf L}_{0}$ has the block form of \erf{LBlock}, it is immediate that $ {\bf T}_{\mathfrak{R}_{0} ,\mathfrak{I}_{0}}=0$ is necessary, as otherwise ${\cal T}$ would map PRE members to outside of the invariant subspace.  This then implies that $ {\bf T}_{\mathfrak{I}_{0} ,\mathfrak{R}_{0}}=0$ as well, due to the unitary/antiunitary nature of ${\cal T}$.  Secondly, as the PRE is proposed to lie entirely in $\mathfrak{I}_{0}$, the effect of ${\cal T}$ on coordinates outside $\mathfrak{I}_{0}$ is irrelevant.  We collect these arguments in the equations
\bqa
&{\bf T}_{\mathfrak{I}_{0} ,\mathfrak{R}_{0}}={\bf T}_{\mathfrak{R}_{0},\mathfrak{I}_{0}}=0,
\label{jointSymm1}\\
&{ \bf T}^{-1}_{\mathfrak{I}_{0}}{\bf L}_{\mathfrak{I}_{0}}{\bf T}_{\mathfrak{I}_{0}}={\bf L}_{\mathfrak{I}_{0}},\label{jointSymm2}\\
&{\bf T}_0\pmb{x}_{{\rm ss}}=\pmb{x}_{{\rm ss}}
\label{jointSymm3}.
\eqa
The orthonormal basis for $\mathfrak{U}_0$ also defines the basis in which we write $\pmb{x}_{{\rm ss}}$. \blk
Note that \erf{jointSymm3} cannot, in general,  be written solely in terms of ${\bf T}_{\mathfrak{I}_{0}}$ because $\pmb{x}_{{\rm ss}}$ may have support on (translated) coordinates outside of $\mathfrak{I}_{0}$.  An example of this is the resonance fluorescence qubit ME, for which the $u$-axis was an $\mathfrak{I}_{0}$ space but $\pmb{x}_{{\rm ss}}$ has support outside of the $x$-axis (which is the translated image of the $u$-axis). Another way of saying this is that $\mathfrak{I}_{0}$ is determined by ${\bf L}_{0}$ but $\pmb{x}_{{\rm ss}}$ is also dependent upon $\pmb{b}$, as per \erf{steadyState}. \blk

It is worth pointing out that, in principle, \erfs{jointSymm1}{jointSymm3} can be satisfied even when both symmetries are not present at the ME level. 
That is, when searching for PREs that simultaneously have invariant subspace symmetry and Wigner symmetry it is not necessary that the Wigner symmetry apply to the entire $\mathfrak{D}\left({ \mathbb{H}}\right)$; all that is relevant is its action upon $\mathfrak{D}_\mathfrak{I}$ (in contrast to both symmetries existing at the ME level).  In other words, it may be the case that ${ \bf T}^{-1}_{\mathfrak{R}_{0}}{\bf L}_{\mathfrak{R}_{0}}{\bf T}_{\mathfrak{R}_{0}}\neq{\bf L}_{\mathfrak{R}_{0}}$.

\subsection{Qubit examples}
\subsubsection{Resonance fluorescence}
\label{RFCombo}
As a first example of the combination of symmetries, let us consider the example of Ref.~\cite{karasik2011many}.  As a reminder, we have the Wigner symmetry ${\bf T}_{0} ={\rm diag}(-1,1,1)$ and invariant subspaces, $\mathfrak{I}_{0}$, being, respectively, the $u$-axis and the $u=0$ great disc. 
It is immediate that  ${\bf T}_{\mathfrak{R}_{0},\mathfrak{I}_{0}}=0$, so that \erf{jointSymm1} holds and it is clear that the steady state is invariant under the action of ${\cal T}$.  The first invariant subspace
has ${\bf L}_{\mathfrak{I}_{0}}=-\gamma/2$ and ${\bf T}_{\mathfrak{I}_{0}}=-1$, both scalars, so that \erf{jointSymm2} holds. 
The second invariant subspace, the great disc, has $u=0$ so that ${\bf T}_{\mathfrak{I}_{0}}$ acts as the identity.   We also have ${\bf T}^{-1}{\bf L}{\bf T}={\bf L}$, so that the Wigner symmetry applies across the entire $\mathfrak{D}\left({ \mathbb{H}}\right)$, not just $\mathfrak{D}_\mathfrak{I}$. Thus, the Wigner and invariant subspace symmetries can co-exist both on a ME level and at that of the PREs.  Indeed a non-trivial manifestation of the Wigner symmetry is found in \frf{2KPREs}(c), which shows a PRE constrained to the Bloch ball image of the \blk first mentioned invariant subspace.

\subsubsection{Absorption and emission ME}
\label{absDisCombo}
Returning to the ME of \erf{absDis}, we will now exploit the abundance of symmetry that it possesses to find PREs of arbitrary size $K$.  We avail ourselves of the maximum amount of  PRE Wigner symmetry; we take it to be of the form $\mathbb{Z}_{K}$, so that 
\beq
{\cal T}\ket{\phi_k}\bra{\phi_k} = \ket{\phi_{{\rm mod}(k+1, K)}}\bra{\phi_{{\rm mod}(k+1, K)}}.
\label{inducedCombo}
\eeq
This means that of the $K$ matrix equations in \erf{jumpCond} only one of them need be considered --- the rest are guaranteed to be satisfied due to the Wigner symmetry.  The explicit form of ${\cal T}$ is an azimuthal rotation of magnitude $\theta=2\pi/K$.  Consequently, each ensemble member must have the same $w$ coordinate --- the entire ensemble lies in the $w=0$ invariant subspace of $\mathfrak{U}_0$. Furthermore, if a single PRE exists, of the nature described, then, due to the Wigner family of PRE symmetry, there must exist an infinity of PREs such that every point of the $w=0$ circle of $\mathfrak{U}_0$ hosts an ensemble member that belongs to atleast one PRE.  This allows us to completely fix all coordinates of the state $\ket{\phi_k}\bra{\phi_k}$ and it can be written as, for example, $\{ \sqrt{1-{\gamma^{2}_{\Delta}/\gamma^{2}_{\Sigma}}},0,0\}^{{\rm T }}$ (any point on the $w=0$ circle could have been chosen).  
Once we have found a single PRE then, of course, the entire family is obtained by rotation.
In summary, the potential PRE that we are investigating consists of $K$ evenly spaced states lying on the $w=0$ circle and, without loss of generality, we take one of them as lying on the $u$-axis. 

Whether this PRE exists or not is determined by \erf{jumpCond}.  We insert the PRE states into these equations and the previously difficult to solve system of polynomial equations is reduced to only two {\it linear} equations that constrain the transition rates away from the $k$th state.  Taking the one state that determines them all to be labelled as $k=1$, we have
\bqa
\sum_{j=2}^{K}\kappa_{j,1}\left(1-\cos\left[\frac{2\pi (j -1) \blk}{K}\right]\right)=\frac{\gamma_{\Sigma}}{2}\label{lin1}
\\
\sum_{j=2}^{K}\kappa_{j,1}\sin\left[\frac{2\pi (j -1)}{K}\right]= 0,\label{lin2}
\eqa
where $\kappa_{j,1}$ denotes the transition rate from state $1$ to state $j$.  The above equations can be satisfied for arbitrary $K\geq 2$.  By setting $K=2$ one regains the PREs described in \srf{invSubAbDis}. It is also clear (in a mathematical sense) that no cyclic $K=3$ PREs can possess the Wigner symmetry.  In fact, from \erf{lin2} and \erf{inducedMap}, no cyclic PREs possess this Wigner symmetry for {\it any} $K>2$.  To find actual existing PREs from \erfs{lin1}{lin2}, non-cyclic PREs need to be considered.  This is because, generally speaking, the two constraints require two free parameters to have a solution.  In this way we {\it can} find PREs having a minimum of two outward transitions per member.  Specifically, for $K=3$, each of the two transition rates is given by $\gamma_{\Sigma}/6$.  A PRE from this family is illustrated in \frf{daryPRE3and4k}(a). 

\begin{figure} [t]
\centering
\includegraphics[width=0.50\textwidth]{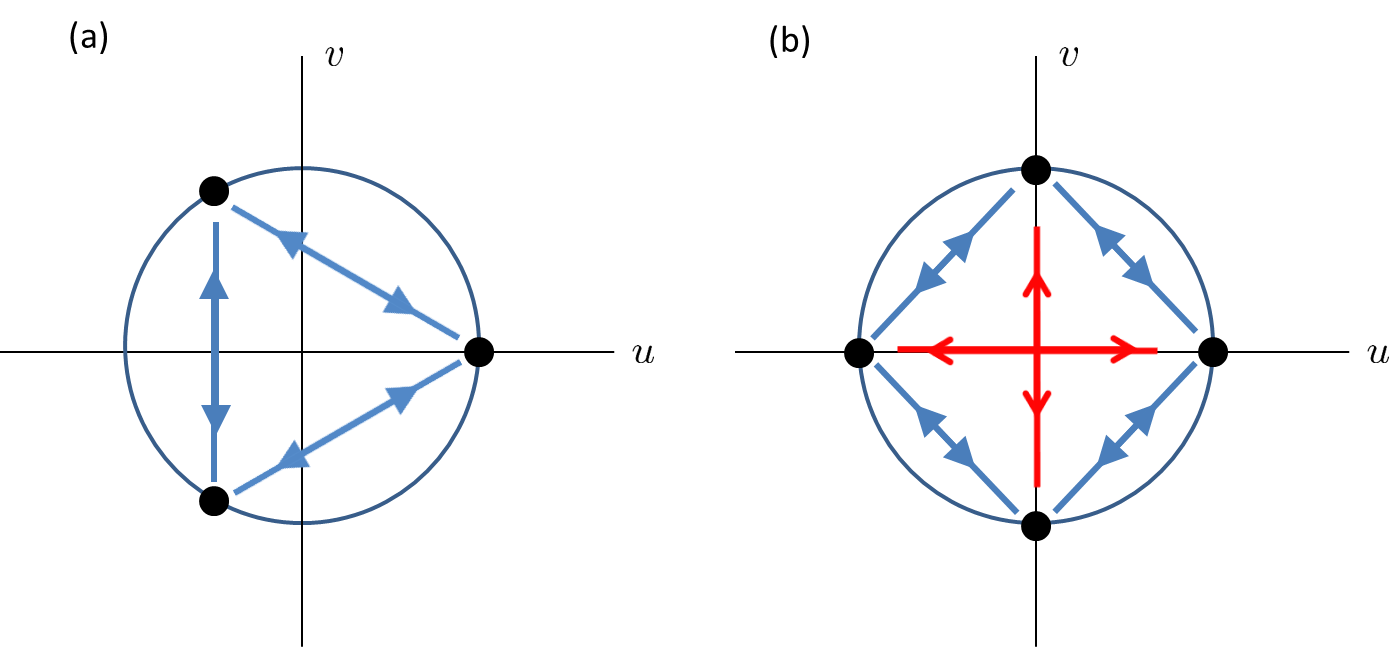} 
\vspace{-2mm}
\caption{In (a) we show a $K=3$ PRE possessing the Wigner symmetry of \erf{inducedCombo}.  As the PRE is entirely contained in the $w=0$ disc, we show this two-dimensional space.  The PRE members are shown as black dots.  All transition rates are of the same magnitude ($\gamma_{\Sigma}/6$) and each member has equal occupation probability.  In (b) we show a $K=4$ PRE, also in the $w=0$ plane.  All the blue transitions are of equal magnitude and are non-zero.  All red transitions are also of equal magnitude but there exists the possibility of them all being zero.  An infinity of PREs can be obtained by arbitrary azimuthal rotation. }
\label{daryPRE3and4k}
\end{figure}

If more than two transitions are allowed (possible when $K>3$), then the excess parameters, as compared to constraints, gives hyperplanes of solutions in $\kappa$ space.  Specifically, for $K=4$, we must have equal transition rates to the two ensemble members that are obtained by a magnitude $\pi/2$ azimuthal rotation ($\kappa_{2,1}=\kappa_{4,1}$).  Additionally, the following constraint must be satisfied $\kappa_{2,1}+\kappa_{3,1}=\gamma_{\Sigma}/4$ (a line of solutions in $\{ \kappa_{2,1},\kappa_{3,1} \}$ space --- note that $\kappa_{2,1}=0$ would return us to the $K=2$ PRE as the $K=4$ PRE would get trapped in a $K=2$ subcycle).  
Additionally, it is of interest that we do have freedom to set some of the transitions to zero for $K=4$: $\kappa_{3,1}=0$, together with $\kappa_{2,1}=\kappa_{4,1}=\gamma_{\Sigma}/4$ satisfies the constraints.  We show a sample $K=4$ PRE in \frf{daryPRE3and4k}(b). \blk
In summary, by using our theory of symmetry to the maximum extent, we have found non-cyclic PREs for arbitrarily large $K$ --- a task that would naively be undertaken by trying to solve a, typically intractable, set of polynomial equations.


\section{Discussion and conclusion}
\label{conc}

In this paper we first discussed the difficulty of finding PREs, then provided some new tools to reduce the complexity of this task.  That is, we have considered symmetries of MEs that allow a simplified search for PREs that comprise a subset (that may be empty, but in many cases we have shown otherwise) of those that exist. 

 The invariant subspace symmetry involved finding a subspace (of the space occupied by density matrices) to which dynamics are constrained, given an appropriate initialisation.  It was then logical to propose searching within this confined subspace for PREs, a task that can be considerably more simple than searching the entire space.  This was highlighted by considering the class of real-valued density matrices and noting that the expected minimum PRE size is halved (for large $D$) when this symmetry is appropriate.  Due to the exponential scaling, in the ensemble size, of the difficulty of finding PREs, this can make previously intractable searches tractable.  Qubit examples of this symmetry were then analyzed in detail, giving new insights into previously known PREs.  For example, it was shown that many of the known $K=3$ resonance fluorescence ME PREs actually possess the symmetry, as do {\it all} of the $K=2$ PREs.

Next we considered MEs possessing a Wigner symmetry (where the Hilbert space inner product is preserved).  This led to the exciting conclusion that new PREs could be generated from existing ones by using the symmetry.  Also, perhaps most importantly, it was shown that if the potential PRE is assumed to possess the Wigner symmetry then a fraction of the constraints governing their existence become redundant. This serves to reduce in size the polynomial system that must be solved to find a PRE.  

This all comes with the proviso that not all PREs can be found in this manner --- there may exist unsymmetric PREs that escape the scope of the symmetry-based searches.  In fact, for the qubit examples, we find several of these PREs; that is, there exist PREs in both the symmetric and unsymmetric categories.

Finally, we investigated a joint application of the invariant subspace and Wigner symmetries.  For the absorption and emission qubit ME, this allowed us to reduce the constraints defining these symmetric PREs to two {\it linear} equations, essentially making trivial the finding of arbitrarily large PREs.  

We note that finding the adaptive measurement scheme necessary to produce a PRE is typically a less demanding computational task than finding the PRE itself, as the system of coupled polynomials that must be solved is smaller.  For this reason, we have focused on PRE symmetries rather than measurement scheme symmetries. However, a brief discussion of the latter is contained in \cref{app:measSymm,app:appB}.

Having demonstrated the practicable and insightful use of the developed symmetry tools, we may now consider new problems to which they can be applied.  In Refs.~\cite{karasik2011many,karasik2011tracking} a number of important open questions regarding PREs are raised.
The first question is: {\it are there MEs for which the minimally sized PRE is larger than $D$?}  The motivation of this query is that if it is answered in the affirmative then open quantum systems can be said to be harder to track than classical systems, as there exist quantum systems that cannot be tracked by a $D$-state finite resettable classical memory (in contrast to classical systems).  Clearly our techniques that find only a subset of potential PREs cannot be used to rule out their existence, but we can, of course, place an upper bound on the ensemble size by finding a PRE using symmetry.
The second question is as follows: {\it is an ensemble size of $K=(D-1)^2+1$ always sufficient for a PRE to be found?}  Once again, our techniques do not allow a direct answer, but would provide an upper bound on the difficulty.
The third question is: {\it does the ensemble size of $K=(D-1)^2+1$ from Ref.~\cite{karasik2011many} reliably predict whether PREs are feasible for a ME of a given form?}  To this question, our techniques and results are directly relevant as we showed that $K=(D-1)^2+1$ is potentially larger than necessary if symmetry is present.  Specifically, \blk we showed that ensembles of only $K=(D^{2}-D+2)/2$ pure states can be expected to be sufficient in the case of real-valued density matrices.  But rather than seeing this 
as an immediate negative answer to the third question, we see it as 
pointing the way to refinement of \blk  the heuristic governing PRE existence in the presence of invariant subspace symmetry. Since we have reduced the PRE size of interest (in the symmetric case) we should be able to test the presaged refined heuristic \blk more easily.  In addition to addressing all the above open questions, \blk we expect to provide further refining of the anticipated PRE size based on other considerations (such as the number of decoherence channels, $L$) in future work~\cite{warWis}.

As a final topic for future work, we suggest that the study of a $D=4$ composite system (two qubits) in terms of the statistics and dynamics of entanglement of discovered PREs would be intriguing.  The construction of multi-qubit systems in a symmetric manner may make possible such an investigation. 

\ack
This work was supported by the Australian Research Council via discovery project number DP130103715, via the Center of Excellence in Engineered Quantum Systems (EQuS), project numbers CE110001013 
and CE170100009, \blk and 
via the Centre for Quantum Computation and Communication Technology (CQC2T), project numbers CE110001027 and CE170100012.  P.W.~is the recipient of a Postgraduate Scholarship 
from the University of Sydney Faculty of Science. H.M.W.~acknowledges the traditional owners of the land on which this work was undertaken at Griffith University, the Yuggera people. 

\appendix
\section{Defective ${\bf L}_{0}$}
\label{notDiagonalizable}
A `defective ${\bf L}_{0}$', is one that is not diagonalizable, which is completely equivalent to that $(D^{2}-1)\times (D^{2}-1)$ matrix not possessing $D^{2}-1$ linearly independent right-eigenvectors~\cite{anton2010elementary}.  The situation of defective ${\bf L}_{0}$ is almost never relevant as nearly all square matrices are diagonalizable over the complex field: those that are not form a set of measure zero and are infinitesimally close to a matrix that is diagonalizable~\cite{golub2012matrix}.  One sufficient (but not necessary) condition guaranteeing diagonalizability, that is of relevance to our work, is that all eigenvalues of ${\bf L}_{0}$ are distinct.  This is the case
for the resonance fluorescence example of \srf{RFa} when $\Omega\neq \gamma/4$.  The other qubit ME that we focus on, the absorption and emission ME of \srf{invSubAbDis}, has a diagonal ${\bf L}_{0}$ by its definition (despite having repeated eigenvalues).  Before coming back to the resonance fluorescence example with $\Omega= \gamma/4$ (which provides an example of defective ${\bf L}_{0}$), we will discuss the non-diagonalizable case more generally.

The procedure for finding invariant subspaces of ${\bf L}_{0}$, when ${\bf L}_{0}$ is not diagonalizable, is very similar to the diagonalizable case; {\it generalized} right-eigenvectors are calculated in the absence of linearly independent ordinary eigenvectors~\cite{strang1993introduction}.  In a canonical basis, these generalized eignevectors are organized into Jordan chains.  The Jordan chain, generated by a generalized eigenvector, $\bar{\pmb{e}}_{m}$, of rank $m$ is comprised of $\left\{\bar{\pmb{e}}_{1},...,\bar{\pmb{e}}_{m} \right\}$ and is defined by
\beq
\bar{\pmb{e}}_{j}=\left({\bf L}_{0}-\lambda\mathbb{1}\right)^{m-j}\bar{\pmb{e}}_{m}
\quad
{\rm for} \, j=1,2,...,m-1,
\eeq 
where $\lambda$ is the defective eigenvalue associated with this particular Jordan chain.  The bar notation is used to differentiate generalized eigenvectors, however, $\bar{\pmb{e}}_{1}=\pmb{e}_{1}$.  That is, the Jordan chain terminates in an ordinary eigenvector, for which $\left({\bf L}_{0}-\lambda\mathbb{1}\right)\bar{\pmb{e}}_{1}=0$.  As per ordinary eigenvectors, the generalised eigenvectors will appear as conjugate pairs, so that real-valued combinations can be formed by taking real and imaginary parts.

To form an invariant subspace, it is no longer sufficient, in general, to take the plane defined by a conjugate pair of generalized eigenvectors.  Under the action of ${\bf L}_{0}$, the state will leak down through the associated Jordan chains.  That is, given an initialization in the space defined by $\bar{\pmb{e}}_{j}$, the state will be confined to the region defined by span $\left\{\bar{\pmb{e}}_{1},...,\bar{\pmb{e}}_{j} \right\}$.  To form an invariant subspace, one can take this region, which becomes larger as $j$ is increased.  The span of different Jordan chains corresponding to different defective eigenvalues can be combined to form even larger invariant subspaces.

For the resonance fluorescence ME, when $\Omega= \gamma/4$, there exist two ordinary eigenvectors $ \pmb{e}_1=(1,0,0)^{{\rm T}}$ and (unnormalized) $ \pmb{e}_{2}=(0,1,1)^{{\rm T}}$.  The latter eigenvector is associated, via a Jordan chain, with a generalized eigenvector of rank 2 that lies in the rebit plane.  Thus, even at this point in parameter space where ${\bf L}_{0}$ is defective, the rebit plane forms an invariant subspace.

\blk

\section{Measurement scheme symmetry --- invariant subspaces}
\label{app:measSymm}
\blk
The symmetries discussed in this paper can be formulated in terms of the measurement scheme that is utilized to realize the PREs.   To do so, we define measurement superoperators, ${\cal O}_{r}$, correspondng to measurement result $r$ (either jump or no-jump evolution), that take the a-priori system state, $\rho$, to the conditioned a-posteriori state, $\rho_{r}$, as per~\cite{WisMil10}
\beq
\tilde{\rho}_{r}={\cal O}_{r}\rho.
\label{unNorm}
\eeq
The tilde indicates an unnormalized state (the norm provides the probability of obtained the result $r$).  We say that the measurement scheme possesses the invariant subspace symmetry iff 
\beq
\forall r,\quad {\rho}_{r}=\frac{{\cal O}_{r}\rho_{\mathfrak{I}}}{\Tr[{\cal O}_{r}\rho_\mathfrak{I}]}
 \in\mathfrak{D}_\mathfrak{I},
 \eeq
for any $\rho_{\mathfrak{I}}\in\mathfrak{D}_\mathfrak{I}$.  If this is satisfied, then the unconditional evolution, described by Lindbladian superoperator, ${\cal L}$, will also preserve $\mathfrak{D}_\mathfrak{I}$, as is required for a consistant definition.

The mathematical structure, developed in \srf{invSubspaceSec}, describing the invariant subspaces of a ME cannot be fully applied to the measurement scheme without modification.  This is because the linearly acting measurement superoperators do not give a normalized state (as indicated in \erf{unNorm}).  It is still possible to represent the unnormalized system state, $\tilde{\rho}$, analogously to \erf{genBloch} (that is, as a weighted sum over generalized Pauli matrices) but now $\dot{r}_{D^2}\neq 0$. It is also true that, in general, ${\cal O}_{r}\rho_{{\rm ss}}\neq 0$, so that we cannot use $\rho_{{\rm ss}}$ as the traceful member of our operator basis in order to reduce the dynamics down to a $D^2-1$ subspace.
The consequence is that it is not sufficient to consider only ${\bf O}^{0}_{r}$ (where ${\bf O}^{0}_{r}$, analogously to ${\bf L}_{0}$, is the restriction of the matrix representation, ${\bf O}_{r}$, of the superoperator ${\cal O}_{r}$ to traceless Hermitian matrices --- a superscript `$0$' is used to avoid a double subscript).  In particular, whether ${\bf O}^{0}_{r}$ has the block form given in \erf{LBlock} is not sufficient to determine if ${\cal O}_{r}$ will preserve the ME's invariant subspace.  Despite this complication, it is of course simple to directly verify whether a previously discovered ME invariant subspace is preserved by ${\cal O}_{r}$.
We now briefly provide two examples for which the measurement scheme alternatively does not and does preserve $\mathfrak{D}_\mathfrak{I}$.

Firstly, consider the $K=2$ PREs of the resonance fluorescence ME described in \erf{LBlockResFlu}.  As explained in Ref.~\cite{karasik2011tracking}, the measurement scheme that realizes the PRE contained in the one-dimensional invariant space attributed to $\pmb{e}_1=(1,0,0)^{{\rm T}}$ consists of switching the sign of a purely imaginary, amplitude $\frac{1}{2}$, LO (remember that, for this example, we take $\sigma_{x}$ as the imaginary Pauli operator) upon a detection event.  (See \erfs{jumpOps}{noJumpOps} for the unravellings as a function of the LO amplitude.)  
It can be shown that for an arbitrary {\it impure} state within this invariant space, a detection event (under the described experimental setup) leads to an increased probability of the ground state being occupied and, thus, the post-jump state is outside the invariant subspace.  The only state belonging to $\mathfrak{D}_\mathfrak{I}$ that remains in $\mathfrak{D}_\mathfrak{I}$ after a jump is the appropriate PRE member.  Similarly, the no-jump evolution also maps impure states belonging to $\mathfrak{D}_\mathfrak{I}$ outside of $\mathfrak{D}_\mathfrak{I}$.  The no-jump evolution increases the excited state occupation probability of such states, as can be inferred from the average evolution preserving $\mathfrak{D}_\mathfrak{I}$  (with the latter being true by our definition of an invariant subspace).

In contrast, to provide an example of a measurement scheme that does possess the invariant subspace symmetry, consider the $K=3$ PREs, arising from $ \pmb{e}_{\pm}$, that are contained in the two-dimensional invariant subspace consisting of the rebit great disc.
The LO amplitude that achieves these $K=3$ PREs is purely real, with the implication that both the jump and no-jump evolution map real-valued density matrices to real-valued density matrices, thus preserving $\mathfrak{D}_\mathfrak{I}$.  Consequently, the measurement scheme possesses the invariant subspace symmetry as we have defined it. 

\section{Measurement scheme symmetry ---Wigner invariance}
\label{app:appB}

The Wigner symmetry of \srf{uniSymm} can also be related to the measurement scheme.  
To do so, we use the form of the jump operator, given in \erf{jumpOps}, to relate pre-jump states $l,l^{\prime}$ and post jump states $k,k^{\prime}$  (with  $l,l^{\prime}$ and $k,k^{\prime}$ respectively related by the induced mapping of \erf{induced}: $P(k)=k^{\prime}$ and $P(l)=l^{\prime}$).  This leads to the following constraint:
\beq
 {\cal T}^{-1}{\cal J}[\hat{c}_{k^{\prime}}^{l^{\prime}}] {\cal T}\ket{\phi_l}\bra{\phi_l}\propto{\cal J}[\hat{c}_{k}^{l}]\ket{\phi_l}\bra{\phi_l},
 \label{consMeas}
 \eeq
%
where we have included the superscript to make explicit that the measurement settings will be dependent upon the current occupied state in the ensemble --- that is, the measurement scheme is adaptive. Additionally, for notational simplicity, we have used the superoperator ${\cal J}[\hat{c}_{k}^{l}]\rho=\hat{c}_{k}^{l}\rho \hat{c}_{k}^{l\dag}$ with
\beq
\hat{c}_{k}^{l}=\left(\sum_{j=1}^{L} S_{kj}^{l} \hat{c}_j + \beta^{l}_{k}\right) .
\eeq
The constraint of \erf{consMeas} must hold, but has dependence upon the states of the PRE.  To define what we mean by the measurement scheme having the Wigner symmetry, we make our definition independent of knowledge of the PRE states, and require that
\beq
 {\cal T}^{-1}{\cal J}[\hat{c}_{k^{\prime}}^{l^{\prime}}] {\cal T}={\cal J}[\hat{c}_{k}^{l}],
 \label{measSymm2}
\eeq
where it is perhaps helpful to keep in mind that matrix representations of the superoperators can be used.

To illustrate this, with the $K=2$ PREs of the resonance fluorescence ME, we can use the known Wigner symmetry, having matrix representation ${\bf T} = {\rm diag}(-1,1,1,1)$, together with the matrix representation of ${\cal J}[\hat{c}_{k}^{l}]$.  Using the oft-discussed properties of this ME, we can establish that only when the LO amplitude is purely imaginary does the measurement scheme possess the symmetry defined in \erf{measSymm2}.  This is the case for the single $K=2$ PRE associated with $\pmb{e}_1=(1,0,0)^{{\rm T}}$, but not the other two $K=2$ PREs.

Note that if \erf{measSymm2} holds then, provided that the reversible and irreversible evolution are separately ${\cal T}$-invariant (which will typically be the case as the parameter describing the Hamiltonian --- $\Omega$ in the case of resonance fluorescence --- will most commonly be independent from that describing the decoherence, $\gamma$ for example), it is simple to show that the no-jump constraints $\hat{H}^{l^{\prime}}_{\rm eff} \ket{\phi_{l^{\prime}}}\propto\ket{\phi_{l^{\prime}}}$ and $\hat{H}^{l}_{\rm eff} \ket{\phi_{l}}\propto\ket{\phi_{l}}$ are automatically satisfied.  The constraint, \erf{measSymm2}, also relates measurement schemes of different PREs in the case that the PREs belong to the same ${\cal T}$ Wigner-symmetric family (see \srf{WIFs}).


\section*{References}
\bibliographystyle{unsrt}
\bibliography{bibliography8}

\begin{thebibliography}{10}

\bibitem{karasik2011many}
R.~I. Karasik and H.~M. Wiseman.
\newblock How many bits does it take to track an open quantum system?
\newblock {\em Phys. Rev. Lett.}, 106(2):020406, 2011.

\bibitem{2001quant.ph..1012H}
L.~{Hardy}.
\newblock {Quantum theory from five reasonable axioms}, January 2001.
\newblock arXiv:quant-ph/0101012.

\bibitem{Gu2012}
M.~Gu, K.~Wiesner, E.~Rieper, and V.~Vedral.
\newblock Quantum mechanics can reduce the complexity of classical models.
\newblock {\em Nat. Commun.}, 3, Mar 2012.

\bibitem{monrasPub}
A.~Monras, A.~Beige, and K.~Wiesner.
\newblock Hidden quantum {M}arkov models and non-adaptive read-out of many-body
  states.
\newblock {\em Appl. Math. and Comp. Sciences}, 3:93, 2011.

\bibitem{vijay2011observation}
R.~Vijay, D.~H. Slichter, and I.~Siddiqi.
\newblock Observation of quantum jumps in a superconducting artificial atom.
\newblock {\em Phys. Rev. Lett.}, 106(11):110502, 2011.

\bibitem{murch2013observing}
K.~W. Murch, S.~J. Weber, C.~Macklin, and I.~Siddiqi.
\newblock Observing single quantum trajectories of a superconducting quantum
  bit.
\newblock {\em Nature}, 502(7470):211, 2013.

\bibitem{PhysRevX.6.041052}
A.~Chantasri, M.~E. Kimchi-Schwartz, N.~Roch, Ir. Siddiqi, and A.~N. Jordan.
\newblock Quantum trajectories and their statistics for remotely entangled
  quantum bits.
\newblock {\em Phys. Rev. X}, 6:041052, Dec 2016.

\bibitem{PhysRevX.6.011002}
P.~Campagne-Ibarcq, P.~Six, L.~Bretheau, A.~Sarlette, M.~Mirrahimi, P.~Rouchon,
  and B.~Huard.
\newblock Observing quantum state diffusion by heterodyne detection of
  fluorescence.
\newblock {\em Phys. Rev. X}, 6:011002, Jan 2016.

\bibitem{minev2018catch}
Z.~K. Minev, S.~O. Mundhada, S.~Shankar, P.~Reinhold,
  R.~Guti{\'e}rrez-J{\'a}uregui, R.~J. Schoelkopf, M.~Mirrahimi, H.~J.
  Carmichael, and M.~H. Devoret.
\newblock To catch and reverse a quantum jump mid-flight.
\newblock {\em arXiv:1803.00545}, 2018.

\bibitem{PhysRevX.9.011004}
A.~Eddins, J.~M. Kreikebaum, D.~M. Toyli, E.~M. Levenson-Falk, A.~Dove, W.~P.
  Livingston, B.~A. Levitan, L.~C.~G. Govia, A.~A. Clerk, and I.~Siddiqi.
\newblock High-efficiency measurement of an artificial atom embedded in a
  parametric amplifier.
\newblock {\em Phys. Rev. X}, 9:011004, Jan 2019.

\bibitem{karasik2011tracking}
R.~I. Karasik and H.~M. Wiseman.
\newblock Tracking an open quantum system using a finite state machine:
  stability analysis.
\newblock {\em Phys. Rev. A}, 84(5):052120, 2011.

\bibitem{wiseman2001inequivalence}
H.~M. Wiseman and J.~A. Vaccaro.
\newblock Inequivalence of pure state ensembles for open quantum systems: the
  preferred ensembles are those that are physically realizable.
\newblock {\em Phys. Rev. Lett.}, 87(24):240402, 2001.

\bibitem{RevModPhys.75.715}
W.~H. Zurek.
\newblock Decoherence, einselection, and the quantum origins of the classical.
\newblock {\em Rev. Mod. Phys.}, 75:715--775, May 2003.

\bibitem{frowis2018macroscopic}
F.~Fr{\"o}wis, P.~Sekatski, W.~D{\"u}r, N.~Gisin, and N.~Sangouard.
\newblock Macroscopic quantum states: Measures, fragility, and implementations.
\newblock {\em Rev. Mod. Phys.}, 90(2):025004, 2018.

\bibitem{courtois2000efficient}
N.~Courtois, A.~Klimov, J.~Patarin, and A.~Shamir.
\newblock Efficient algorithms for solving overdefined systems of multivariate
  polynomial equations.
\newblock In {\em Advances in Cryptology—EUROCRYPT 2000}, pages 392--407.
  Springer, 2000.

\bibitem{4639467}
F.~Ticozzi and L.~Viola.
\newblock Quantum {M}arkovian subsystems: invariance, attractivity, and
  control.
\newblock {\em IEEE Transactions on Automatic Control}, 53(9):2048--2063, Oct
  2008.

\bibitem{wigner1959group}
E.~Wigner.
\newblock {\em Group theory and its application to the quantum mechanics of
  atomic spectra}.
\newblock Academic Press, New York, 1959.

\bibitem{WisMil10}
H.~M. Wiseman and G.~J. Milburn.
\newblock {\em Quantum measurement and control}.
\newblock Cambridge University Press, 2010.

\bibitem{doi:10.1063/1.522979}
V.~Gorini, A.~Kossakowski, and E.~C.~G. Sudarshan.
\newblock Completely positive dynamical semigroups of {N}-level systems.
\newblock {\em J. Math. Phys.}, 17(5):821--825, 1976.

\bibitem{PhysRevA.49.2133}
H.~M. Wiseman.
\newblock Quantum theory of continuous feedback.
\newblock {\em Phys. Rev. A}, 49:2133--2150, Mar 1994.

\bibitem{CarQTraj}
H.~Carmichael.
\newblock {\em An open systems approach to quantum optics: lectures presented
  at the Universit{\'e} Libre de Bruxelles, October 28 to November 4, 1991},
  volume~18.
\newblock Springer Science \& Business Media, 2009.

\bibitem{wiseman1999quantum}
H.~M. Wiseman and G.~E. Toombes.
\newblock Quantum jumps in a two-level atom: simple theories versus quantum
  trajectories.
\newblock {\em Phys. Rev. A}, 60(3):2474, 1999.

\bibitem{Dol73}
S.~J. Dolinar.
\newblock An optimum receiver for the binary coherent state quantum channel.
\newblock {\em MIT Research Laboratory of Electronics Quarterly Progress
  Report}, 111:115, 1973.

\bibitem{PhysRevLett.75.4587}
H.~M. Wiseman.
\newblock Adaptive phase measurements of optical modes: going beyond the
  marginal \protect{$Q$} distribution.
\newblock {\em Phys. Rev. Lett.}, 75:4587--4590, Dec 1995.

\bibitem{BerHalWis13}
D.~Berry, M.~Hall, and H.~M. Wiseman.
\newblock Stochastic {H}eisenberg limit: optimal estimation of a fluctuating
  phase.
\newblock {\em Phys. Rev. Lett.}, 111:113601, Sep 2013.

\bibitem{PhysRevLett.89.133602}
M.~A. Armen, J.~K. Au, J.~K. Stockton, A.~C. Doherty, and H.~Mabuchi.
\newblock Adaptive homodyne measurement of optical phase.
\newblock {\em Phys. Rev. Lett.}, 89:133602, Sep 2002.

\bibitem{PhysRevLett.104.093601}
T.~A. Wheatley, D.~W. Berry, H.~Yonezawa, D.~Nakane, H.~Arao, D.~T. Pope, T.~C.
  Ralph, H.~M. Wiseman, A.~Furusawa, and E.~H. Huntington.
\newblock Adaptive optical phase estimation using time-symmetric quantum
  smoothing.
\newblock {\em Phys. Rev. Lett.}, 104:093601, Mar 2010.

\bibitem{YH&12}
H.~Yonezawa, D.~Nakane, T.~A. Wheatley, K.~Iwasawa, S.~Takeda, H.~Arao,
  K.~Ohki, K.~Tsumura, D.~W. Berry, T.~C. Ralph, H.~M. Wiseman, E.~H.
  Huntington, and A.~Furusawa.
\newblock Quantum-enhanced optical-phase tracking.
\newblock {\em Science}, 337(6101):1514--1517, 2012.

\bibitem{KIMURA2003339}
G.~Kimura.
\newblock The {B}loch vector for {N}-level systems.
\newblock {\em Phys. Lett. A}, 314(5):339 -- 349, 2003.

\bibitem{PhysRevA.81.062306}
S.~G. Schirmer and X.~Wang.
\newblock Stabilizing open quantum systems by {M}arkovian reservoir
  engineering.
\newblock {\em Phys. Rev. A}, 81:062306, Jun 2010.

\bibitem{1751-8121-41-23-235303}
R.~A. Bertlmann and P.~Krammer.
\newblock Bloch vectors for qudits.
\newblock {\em J. Phys. A}, 41(23):235303, 2008.

\bibitem{bosma1997magma}
W.~Bosma, J.~Cannon, and C.~Playoust.
\newblock The {MAGMA} algebra system i: The user language.
\newblock {\em J. Symb. Comput.}, 24(3):235--265, 1997.

\bibitem{JOHANSSON20131234}
J.~R. Johansson, P.~D. Nation, and F.~Nori.
\newblock Qutip 2: A {P}ython framework for the dynamics of open quantum
  systems.
\newblock {\em Comput. Phys. Commun.}, 184(4):1234 -- 1240, 2013.

\bibitem{faugere1999new}
J.~Faugere.
\newblock A new efficient algorithm for computing {G}r{\"o}bner bases (f 4).
\newblock {\em J. Pure Appl. Algebr.}, 139(1):61--88, 1999.

\bibitem{buchberger1976theoretical}
B.~Buchberger.
\newblock A theoretical basis for the reduction of polynomials to canonical
  forms.
\newblock {\em ACM SIGSAM Bulletin}, 10(3):19--29, 1976.

\bibitem{MAGMAtiming}
{MAGMA} {G}r{\"o}bner basis timings.
\newblock \url{http://magma.maths.usyd.edu.au/~allan/gb/}.
\newblock Accessed: 2017-07-07.

\bibitem{faugere2001finding}
J.~Faug{\`e}re.
\newblock Finding all the solutions of cyclic 9 using {G}r{\"o}bner basis
  techniques.
\newblock {\em Computer Mathematics (Matsuyama, 2001)}, pages 1--12, 2001.

\bibitem{privateSteel}
Private communication with Dr. Allan Steel.

\bibitem{anton2010elementary}
H.~Anton and C.~Rorres.
\newblock {\em Elementary linear algebra: applications version}.
\newblock John Wiley \& Sons, 2010.

\bibitem{warWis}
P.~Warszawski and H.~M. Wiseman.
\newblock {I}n preparation.

\bibitem{1367-2630-16-6-063028}
S.~Daryanoosh and H.~M. Wiseman.
\newblock Quantum jumps are more quantum than quantum diffusion.
\newblock {\em New J. Phys.}, 16(6):063028, 2014.

\bibitem{PhysRevLett.108.170501}
M.~F. Santos, M.~Terra~Cunha, R.~Chaves, and A.~R.~R. Carvalho.
\newblock Quantum computing with incoherent resources and quantum jumps.
\newblock {\em Phys. Rev. Lett.}, 108:170501, Apr 2012.

\bibitem{cox2006using}
D.~A. Cox, J.~Little, and D.~O'Shea.
\newblock {\em Using algebraic geometry}, volume 185.
\newblock Springer Science \& Business Media, 2006.

\bibitem{wigner1}
E.~P. Wigner.
\newblock {\em Gruppentheorie}.
\newblock Vieweg, 1931 pp 251-254.

\bibitem{golub2012matrix}
G.~H. Golub and C.~F. Van~Loan.
\newblock {\em Matrix computations}, volume~3.
\newblock JHU press, 2012.

\bibitem{strang1993introduction}
G.~Strang.
\newblock {\em Introduction to linear algebra}, volume~3.
\newblock Wellesley-Cambridge Press Wellesley, MA, 1993.

\end{thebibliography}

\end{document}